%Paper: hep-ph/9209210
%From: preskill@theory3.caltech.edu (John Preskill)
%Date: Thu, 3 Sep 92 10:03:46 PDT

\input harvmac

\Title{\vbox{\baselineskip12pt\hbox{HUTP-92/A018}\hbox{CALT-68-1786}
\hbox{hep-ph/9209210}}}{Decay of
Metastable Topological Defects}

%For more complicated situations, substitute for {\it either\/} argument:
%\Title{\vbox{\baselineskip12pt\hbox{HUTP-88/A000}\hbox{SLAC-PUB 88-001}
%               \hbox{photocopy at own risk}}}
%{\vbox{\centerline{This title is too long to fit}
%       \vskip2pt\centerline{comfortably on one line*}}}
%   \footnote{}{*optional footnote on title}
%if too many authors for abstract on same page, say   \vfill\eject\pageno0

\centerline{John Preskill}
\smallskip
\centerline{California Institute of Technology}
\centerline{Pasadena, CA \ 91125}

\bigskip

\centerline{and}

\bigskip

\centerline{Alexander Vilenkin\footnote*{On leave from Tufts University,
Medford, MA \ 02155}}
\smallskip
\centerline{California Institute of Technology}
\centerline{Pasadena, CA \ 91125}
\centerline{and}
\centerline{Lyman Laboratory of Physics}
\centerline{Harvard University}
\centerline{Cambridge, MA \ 02138}
\vskip .3in
We systematically analyze the decay of metastable topological defects that
arise from the spontaneous breakdown of gauge or global symmetries.
Quantum--mechanical tunneling rates are estimated for a variety of decay
processes.  The decay rate for a global string, vortex, domain wall, or kink is
typically suppressed compared to the decay rate for its gauged counterpart.  We
also discuss the decay of global texture, and of semilocal and electroweak
strings.
\bigskip
\Date{August, 1992} %replace this line by \draft  for preliminary versions
             %or specify \draftmode at some point
%\draftmode
%if you want double-space, use e.g. \baselineskip=20pt plus 2pt minus 2pt

\vfill\eject
\newsec{Introduction}

Topological defects arise as stable solutions of classical field equations
in a variety of models with spontaneously broken symmetries. The higher
symmetry usually characterizes the high--temperature phase of the model,
and the symmetry breaking corresponds to a phase transition. The type of
defects formed at a phase transition depends on the topology of the vacuum
manifold, $M=G/H$, where $G$ and $H$ are, respectively, the symmetry groups
before and after the symmetry breaking \nref\coleman{S. Coleman, ``Classical
Lumps and Their Quantum Descendants,'' in {\it New Phenomena in Subnuclear
Physics}, ed. A. Zichichi (Plenum, London, 1977).}\nref\kibble{T. W. B. Kibble,
{\it J. Phys.} {\bf A9} (1976) 1387.}\nref\vilenkin{A. Vilenkin,
{\it Phys. Rep.} {\bf 121} (1985) 263.}\nref\preskill{J. Preskill,
``Vortices and
Monopoles,'' in {\it Architecture of the Fundamental Interactions at Short
Distances}, ed. P. Ramond and R. Stora (North-Holland, Amsterdam,
1987).}\refs{\coleman{--}\preskill}.

Linear defects, or strings, are formed if the first homotopy group is
nontrivial, $\pi_1(M) \neq I$; point defects, or monopoles, are formed if
$\pi_2 (M) \neq I$, and sheet--like defects, or domain walls, are formed if
$\pi_0(M) \neq I$. These defects are stable in the sense that ``unwinding'' the
topological knots associated with the defects would require going over an
infinitely high potential barrier. The physical properties of defects
crucially depend on whether the broken symmetry is gauge or global.
For example, the mass of a global monopole and the mass per unit length of a
global string are infrared--divergent, while the corresponding quantities
for gauge defects are finite.

The purpose of this paper is to give a systematic account of the decay of {\it
metastable} defects in relativistic field theories.  A metastable defect is a
stable solution to the classical
field equations, stable in the sense that small vibrations about the solution
have non-negative frequency squared.  But a metastable defect can be unwound by
going over a {\it finite} potential barrier; hence, it can decay quantum
mechanically.  In the limit of small $\hbar$, the decay rate approaches zero
like $e^{-B/\hbar}$.  We will describe how the ``tunneling action'' $B$ can be
calculated for various types of metastable defects.

One type of metastable defect can arise
in models with a sequence of phase transitions,
\eqn\eIi{G \rightarrow H_1 \rightarrow H_2.}
\noindent Defects will be formed if the manifolds $M_1 = G/H_1$ and $M_2 =
H_1/H_2$
have nontrivial homotopy groups. However, these defects will not be
topologically stable if $M= G/H_2$ has trivial topology. Consider, for
example, the sequence
\eqn\eIii{G \rightarrow Z_2 \rightarrow I,}
\noindent with $\pi_0(G) = \pi_1 (G) = I$. Since $\pi_1(G/Z_2) = Z_2$ and
$\pi_0 (Z_2) = Z_2$, the first phase transition gives rise to strings and
the second to domain walls. However, it can be shown \nref\kls{T. W. B.
Kibble,
G. Lazarides, and Q. Shafi, {\it Phys. Rev.} {\bf D26} (1982)
435.}\nref\evervil{A. E. Everett and A. Vilenkin, {\it Nucl. Phys.} {\bf
B207} (1982) 43.}\refs{\kls,\evervil} that
strings
formed at the first phase transition become boundaries of the walls formed
at the second phase transition. Closed and infinite walls without
boundaries can also be formed, but they are not topologically stable: an
infinite planar wall decays by spontaneous nucleation of circular holes
bounded by strings. Quite similarly, the sequence
\eqn\eIiii{G \rightarrow U(1) \rightarrow I,}
\noindent with $\pi_2(G) = \pi_1 (G) = I$ leads to formation of monopoles
which get connected by strings. The strings formed at the second phase
transition are metastable and decay by nucleation of
monopole--antimonopole pairs.

These observations are not new.  Indeed,
the decay rate of a metastable string was estimated in
Ref. \ref\vildecay{A. Vilenkin, {\it Nucl. Phys.} {\bf B196} (1982) 240.},
and the decay rate of a metastable wall was estimated in Ref. \kls,
assuming that the thickness of the defects can be neglected.
Our intent is to discuss such tunneling phenomena in a fairly comprehensive
way.  In various cases, we describe the instanton (or ``bounce'') corresponding
to the decay, and estimate the tunneling action.
While some of our calculations  merely rederive familiar results, we also
present a number of new results.  We  consider defects in
$D=1,\; 2,$ and $3$ spatial dimensions, and  discuss the connection between
tunneling phenomena in different dimensions.   (Some of these
lower--dimensional defects may have applications to condensed matter physics.)
We emphasize in particular the differences between defects arising from gauge
and global symmetries.  Because global defects have long-range interactions
mediated by massless Nambu--Goldstone bosons,  the decay of a global defect is
typically suppressed compared to the decay of its gauged counterpart.

There are also other types of metastable defects that are not associated
with a hierarchy of symmetry breakdown of the form eq.~\eIi.  One interesting
example is global texture \kibble, and we discuss the decay of metastable
texture in
various dimensions.  Another interesting case is the electroweak string
\ref\electroweak{T. Vachaspati, ``Electroweak Strings,'' Tufts Preprint
TUTP-92-3 (1992); {\it Phys. Rev. Lett.} {\bf 68} (1992) 1977.} (or
vortex), which might occur in realistic extensions of the standard model; we
analyze its decay as well.

We outline a general classification of metastable defects in Section 2, and
then proceed in the remainder of the paper to discuss various special cases in
more detail.  Section 3 concerns defects that arise from a hierarchy of gauge
symmetry breaking---monopoles, strings, domain walls, and their lower
dimensional analogs.  Section 4 analyzes the consequences of a hierarchy of
global symmetry breaking.  Metastable defects arising from the intrinsic
breaking of a spontaneously broken global symmetry by a small perturbation
(like axion domain walls) are discussed in Section 5.  We consider in Section 6
heavy metastable defects that decay to light stable defects.  The decay of
global texture is treated in Section 7. Electroweak and ``semilocal'' defects
are discussed in Section 8.  Section 9 contains our conclusions,
including some remarks about the cosmological implications of
metastable defects.

\newsec{General Theory}
In this section, we formulate the general theory of metastable topological
defects.  This theory will be illuminated later by various examples.

The metastable defects that we will discuss fall into three broad categories.
Those in the first category are associated with a hierarchy of (gauge or
global) symmetry breakdown.  Those in the second category are associated with
the intrinsic breaking of a global symmetry by a small perturbation.  Those in
the third category do not fit into either of the first two categories---they
are classically stable but are not prevented from decaying by any topological
conservation law.  (Examples include global texture and electroweak vortices.)

\medskip

\noindent {\bf a) Hierarchy of Symmetry Breakdown}

Let us first consider models with a sequence of phase transitions
\eqn\aIIi{G\to H_1\to H_2.}
Here G is a finite-dimensional compact Lie group that we may take to be
connected.  It may be either a global symmetry group or a gauge group.  (The
distinction between global and gauge symmetry will be discussed later.)  The
$G$ symmetry breaks to the subgroup $H_1$ at the mass scale $\eta_1$, and then
breaks further to $H_2\subset H_1$ at the much lower mass scale $\eta_2$.  We
wish to address whether topological defects associated with the second stage of
symmetry breakdown remain topologically stable when $H_1$ is embedded in $G$,
and also the closely related but somewhat different question whether
topological defects produced in the first stage survive when the second
symmetry breakdown occurs.

\medskip

\noindent {\it i) Codimension 1}

By a codimension 1 defect we mean one of dimension $D-1$ in $D$ spatial
dimensions---it is a domain wall, or, in $D=1$, a particle or ``kink.''
Topologically stable codimension 1 defects exist if the vacuum manifold is
disconnected.  Thus, if the symmetry group $H_1$ breaks to $H_2$ (and assuming
no accidental degeneracy), these defects are classified by the homotopy group
$\pi_0(H_1/H_2)$.  But if $H_1$ is actually embedded in a larger symmetry group
$G$ that breaks at a much larger mass scale, then this defect may not be
absolutely stable (although it is very long-lived).  The domain wall separates
two regions in which the order parameter takes values in two distinct connected
components of $H_1/H_2$.  If these components are connected in the larger
manifold $G/H_2$, then the domain wall is metastable.  Mathematically, since
$H_1/H_2$ is included in $G/H_2$, there is a natural homomorphism
\eqn\aIIii{\pi_0(H_1/H_2)\longrightarrow\pi_0(G/H_2).}
Metastable defects of codimension 1 are classified by the nontrivial elements
of the kernel of this homomorphism.

Associated with each nontrivial element of this kernel, there is a path in $G$
that begins at the identity and ends at an element of $H_1$ that is not
connected to the identity in $H_1$.  This path defines a representative of a
nontrivial element homotopy class in $\pi_1(G/H_1)$.  Associated with this
class there is a string or vortex  that arises in the symmetry breakdown $G\to
H_1$.  The physical interpretation is that the metastable codimension 1 defect
can end on a codimension 2 defect.

\medskip

\noindent {\it ii) Codimension 2}

A codimension 2 defect is a ``string'', or, in D=2, a particle or ``vortex.''
By reasoning analogous to that above, metastable defects of codimension 2 are
classified by the nontrivial elements of the kernel of the homomorphism
\eqn\aIIiii{\pi_1(H_1/H_2)\longrightarrow\pi_1(G/H_2).}
Associated with each nontrivial element of the kernel, there is a
noncontractible closed loop in $H_1$ that can be deformed to a point in $G$.
This deformation of the loop defines a nontrivial element of
\eqn\aIIiv{\pi_2(G/H_1)=\pi_1(H_1)/\pi_1(G).}
Associated with this element is a magnetic monopole that arises when $G$ breaks
to $H_1$.
The physical interpretation is that the metastable codimension 2 defect can end
on a codimension 3 defect.

\medskip

\noindent {\it iii) Codimension 3}

A codimension 3 defect is a ``monopole,'' a particle in $D=3$.  In principle,
metastable monopoles are classified by the nontrivial elements of the kernel of
the homomorphism
\eqn\aIIv{\pi_2(H_1/H_2)\longrightarrow\pi_2(G/H_2).}
However, this kernel is always trivial---metastable monopoles do not exist.
Mathematically, this is because $\pi_2(H_1)=I$ for any finite-dimensional
compact Lie group $H_1$.  (Note that metastable domain walls are associated
with nontrivial elements of $\pi_0(H_1)$, and metastable strings are associated
with nontrivial elements of $\pi_1(H_1)$.)

We can express this result in more physical terms in the case where $G$ is a
gauge symmetry.  Then
the magnetic
monopole that arises when $H_1$ breaks to $H_2$ carries a conserved magnetic
charge, which can be detected by measuring the long-range $H_2$ magnetic
field of the monopole.  Embedding
$H_1$ in $G$ does not extinguish that long-range field, or destroy the
conservation law---the monopole remains absolutely stable.  (Although quantum
effects, specifically  color confinement, may cause the magnetic field to be
screened, these effects do not prevent the charge from being detected at long
range, and do not destroy the conservation law \nref\colemanother{S. Coleman,
``The Magnetic Monopole Fifty Years Later,'' in {\it The Unity of the
Fundamental Interactions}, ed. A. Zichichi (Plenum, New York,
1983).}\nref\susskind{M. Srednicki and L. Susskind, {\it Nucl. Phys.} {\bf
B179} (1981) 239.}\refs{\colemanother,\susskind}.)

\medskip

\noindent {\bf b) Comments}

\medskip

\noindent {\it i)  Bianchi Identity}

We could also consider a more intricate symmetry breaking hierarchy of the form
\eqn\aIIvi{G\to H_1\to H_2\to H_3.}
One might then wonder if it is possible for a monopole to arise when $G$ breaks
to $H_1$ such that the monopole becomes attached to a string when $H_1$ breaks
to $H_2$, and the string in turn becomes attached to a wall when $H_2$ breaks
to $H_3$.  It is easy to see that this is not possible.  This conclusion is
probably best understood as a consequence of the Bianchi identity---``the
boundary of a boundary is zero.''  If a string is the boundary of a domain
wall, then the string cannot end (on a monopole).  In terms of the above
homotopy classification, we saw that there are two types of strings that can
arise when $H_1$ breaks to $H_2$.  A string that ends on a monopole is
associated with a noncontractible closed path in $H_1$, while a string that
bounds a domain wall is associated with an {\it open} path in $H_1$ that begins
at  the identity and ends at an element of $H_2$ that is not connected to the
identity in $H_2$.

Another observation is closely related to the above:  It is impossible for a
string that ends on a monopole to have nontrivial Aharonov-Bohm interactions
that can detect the ``quantum hair'' \ref\krauss{L. M. Krauss and F. Wilczek,
{\it Phys. Rev. Lett.} {\bf 62} (1989) 1221.} of charged particles.  Strings
that detect
quantum hair carry a magnetic flux that does not lie in the connected component
of the unbroken group $H_2$; they are the kind of strings
that can bound domain walls, not the kind that can end on monopoles
\nref\gupta{A. K. Gupta, J. Hughes, J. Preskill, and M. B. Wise, {\it Nucl.
Phys.} {\bf B333} (1990) 195.}\nref\prekra{J. Preskill and L. M. Krauss, {\it
Nucl. Phys.} {\bf B341} (1990) 50.}\refs{\gupta,\prekra}.  This is
not to say that the Aharonov--Bohm interactions of strings that end on
monopoles must be completely trivial; rather, the group element that
characterizes the flux trapped in the core of the string must be connected to
the identity in $H_2$.  An example of a string that ends on a monopole, yet has
nontrivial Aharonov--Bohm interactions, is the electroweak string that we will
discuss in Section 8 \nref\pressemi{J. Preskill, ``Semilocal Defects,'' Caltech
Preprint CALT-68-1787 (1992).}\refs{\electroweak,\pressemi}.

\medskip

\noindent {\it ii)  Bundles}

The analysis in (a) above can be reexpressed in the language of fiber bundles.
When the symmetry breaking pattern \aIIi\ occurs, we may view the vacuum
manifold $G/H_2$ as the total space of a bundle with basespace $G/H_1$, fiber
$H_1/H_2$, and structure group $H_1$.  Then the topological defects arising in
the first stage of the symmetry breakdown are determined by the topology of the
basespace, and the defects arising in the second stage are determined by the
topology of the fiber.  Our criterion for a codimension $n+1$ defect to be
metastable, then, is that a mapping that represents a nontrivial element of
$\pi_n$ of the fiber is topologically trivial in the total space of the bundle.

\medskip

\noindent {\it iii)  Survival}

We may also ask a slightly different question than that formulated in (a)
above.  If a defect arises when $G$ breaks to $H_1$, will that defect
``survive'' if the symmetry breaks further, to $H_2$?  Before, we found the
criterion for a monopole to become attached to a single string, or for a string
to become attached to a single wall.  Now we are asking a more general
question, because it is also possible for a monopole to become attached to more
than one string, or for a string to become attached to more than one wall.

The criterion for a defect to survive is most simply stated in the fiber bundle
language.  A codimension $n+1$ topological defect that arises when $G$ breaks
to $H_1$ is associated with a nontrivial element of $\pi_n$ of the basespace of
the bundle.  The defect survives if this element can be ``lifted'' to $\pi_n$
of the total space of the bundle.  That is, the bundle comes equipped with a
projection map $\phi:~G/H_2\to G/H_1$, and the defect is characterized by a
topologically nontrivial map $f:~S^n\to G/H_1$.  The defect survives if there
is a continuous map $\tilde f:~S^n\to G/H_2$ such that $f=\phi\circ\tilde f$.

Domain walls always survive, but strings and monopoles need not.  Note that it
is possible for a monopole to be attached to two (or more) strings where one
string is
heavy and the other is light.  In $D=2$, then, the monopole mediates the decay
of a heavy vortex to a light vortex (or several light vortices).
If two degenerate strings end on a monopole \ref\hindkib{M. Hindmarsh and T. W.
B. Kibble, {\it Phys. Rev. Lett.} {\bf 55} (1985) 2398.}, then in $D=2$ the
monopole
is an instanton that allows two degenerate vortices to mix quantum
mechanically.
Similarly, a string could be attached to a heavy wall and a light wall.  Then,
in
$D=1$, the vortex mediates the decay of a heavy kink to a light kink.  If two
degenerate walls end on a string, then in $D=1$ the vortex allows degenerate
kinks to mix.

\medskip

\noindent {\bf c)  Intrinsic Symmetry Breaking}

Another type of metastable defect can arise when an {\it approximate} global
symmetry is spontaneously broken.  Consider the pattern
\eqn\aIIvii{\matrix{G_{\rm approx}&\rightarrow&H_{\rm approx}\cr
\cup&&\cup\cr
G_{\rm exact}&\rightarrow& H_{\rm exact}}.}
That is, $G_{\rm approx}$ is spontaneously broken to $H_{\rm approx}$, and is
also intrinsically broken by a small perturbation to $G_{\rm exact}$.  (Thus
$G_{\rm approx}$ must be a global symmetry group; gauge symmetries are always
exact.)  The unbroken exact symmetry is $H_{\rm exact}$, the intersection of
$G_{\rm exact}$ and $H_{\rm approx}$.

If we ignore the intrinsic symmetry breaking, then topological defects of
codimension $n+1$ are classified by $\pi_n(G_{\rm approx}/H_{\rm approx})$.  We
may ask if such a defect can ``survive'' when the intrinsic symmetry breakdown
is taken into account.  The criterion for survival can be expressed in the
following way.  The defect is associated with a topologically nontrivial map
from $S^n$ to the approximate vacuum manifold $G_{\rm approx}/H_{\rm approx}$
The defect survives if this mapping can be continuously deformed to one that
takes values in the exact vacuum manifold $G_{\rm exact}/H_{\rm exact}$.

If a $G_{\rm approx}/H_{\rm approx}$ domain wall does not survive, then the
energy
densities on the two sides of the wall are unequal, and a resulting pressure
pushes the wall away.  If a $G_{\rm approx}/H_{\rm approx}$ string does not
survive, then it becomes attached to one or more walls; if the number of walls
is exactly one, then there is a metastable wall that can end on a loop of
string.  If a $G_{\rm approx}/H_{\rm approx}$  (global) monopole does not
survive,
then it becomes attached to one or more strings; if the number of strings is
exactly one, there is a metastable string that can break by nucleating a
monopole pair.

The most familiar example of this
phenomenon is the axion string, which becomes attached to $N$ axion domain
walls \nref\sikivie{P. Sikivie, {\it Phys. Rev. Lett.} {\bf 48} (1982)
1156.}\nref\lazshaf{G. Lazarides and Q. Shafi, {\it Phys. Lett.} {\bf B115}
(1982) 21.}\nref\everett{A. Vilenkin and A. E. Everett, {\it Phys. Rev. Lett.}
{\bf 48} (1982) 1867.}\refs{\sikivie{--}\everett}.  Thus, if $N=1$, an axion
domain wall is metastable and can decay by
nucleating a loop of axion string.
In $D=1$,
the axion vortex mediates the decay of an axion kink.

\medskip

\noindent {\bf d)  Other Cases}

There are a few interesting classes of metastable defects that do not arise due
to a hierarchy of symmetry breakdown, or because of intrinsic symmetry
breaking.  These defects are classically stable, but are not forbidden to decay
quantum mechanically.

\medskip

\noindent {\it i)  Global Texture}

If a global symmetry $G$ is spontaneously broken to $H$, then a global texture
(or ``skyrmion'') is a field configuration that takes values in (or near) the
vacuum
manifold $G/H$ everywhere.  (There is no ``restoration'' of the spontaneously
broken symmetry in its core.)  Such configurations, if they have finite energy,
are classified in D spatial dimensions by $\pi_D(G/H)$ \kibble.  A texture has
only
gradient energy, and for $D\ge 3$, the gradient energy makes it want to
collapse.  But it can be stabilized if suitable higher derivative terms
\ref\skyrme{T. H. R. Skyrme, {\it Proc. Roy. Soc.} {\bf A260} (1961) 127; E.
Witten, {\it Nucl. Phys.} {\bf B223} (1983) 433.} are
introduced into the action (or if a {\it subgroup} of $G$ is gauged
\ref\hindtex{M. Hindmarsh, ``Semilocal Topological Defects,'' DAMTP Preprint
DAMTP-HEP-92-24 (1992).}).

For $D=1$, its gradient energy makes a texture want to spread out.  It can be
stabilized if space is compactified to a circle of finite circumference.

Unlike a kink in one dimension, a (gauge) vortex in two dimensions, or a
(gauge) monopole in three dimensions, a global texture in D dimensions is not
separated from the vacuum by an infinite energy barrier.  Thus, even if it is
classically stable, there is no topological conservation law that prevents it
from decaying quantum mechanically.

Indeed, in any model that contains a $D$-dimensional global texture, there is
also a ``global instanton'' that mediates the decay of the texture.  This
instanton is also classified by $\pi_D(G/H)$; it is a pointlike defect in
$D+1$-dimensional Euclidean spacetime, with the world line of a texture ending
on the instanton.  We will discuss texture decay in more detail in Section 7.

\medskip

\noindent {\it ii)  Semilocal and Electroweak Strings}

``Semilocal'' strings \ref\vachsemi{T. Vachaspati and A. Ach\'ucarro, {\it
Phys. Rev.} {\bf D44} (1991) 3067.} (or vortices) can arise in models that have
both gauge
and global symmetries that are spontaneously broken, but only if the symmetries
``mix;''  that is, there must be unbroken global symmetry generators that are
nontrivial linear combinations of spontaneously broken gauge symmetry
generators and global symmetry generators.

Consider the pattern
\eqn\aIIviii{\matrix{G_1\times G_2&\rightarrow&H\cr
\cup&&\cup\cr
G_1&\rightarrow& H_1}.}
Here $G_1$ is the gauge group and $G_2$ is a global symmetry group.  $H_1$ is
the unbroken gauge group, the intersection of $G_1$ and $H$.  In this scheme,
for $D=2$, there is a topologically conserved magnetic flux that is classified
by $\pi_1(G_1/H_1)$, and a natural homomorphism
\eqn\aIIix{\pi_1(G_1/H_1)\rightarrow\pi_1([G_1\times G_2]/H).}
If this homomorphism has a nontrivial kernel (which is possible only if $G_1$
and $G_2$ mix \pressemi), then there are field configurations that carry
nontrivial
$G_1/H_1$ magnetic flux, where the order parameter takes values in the vacuum
manifold $[G_1\times G_2]/H$ everywhere.

When such configurations exist, it becomes a dynamical question whether the
energy in a given magnetic flux sector is minimized by a localized vortex or by
a configuration in which the magnetic flux is spread out over an arbitrarily
large area.  The answer depends on the details of the Higgs potential.

In some models, there may be vortices that are classically stable, but are
kinematically allowed to decay to configurations in which the magnetic flux is
spread out.  Then the decay is mediated by a global monopole \nref\hindprl{M.
Hindmarsh, {\it Phys. Rev. Lett.} {\bf 68} (1992)
1263.}\refs{\hindprl,\hindtex,\pressemi}, as we will
describe in more detail in Section 8.  Similarly, for $D=3$, there may be
string that
can decay via the nucleation of a global monopole pair.

The simplest example of the semilocal phenomenon is the case $G_1=U(1)$,
$G_2=SU(2)$, and $H=U(1)$, which may be regarded as the minimal standard
electroweak model in the limit $\sin^2\theta_W=1$.

It is also interesting to consider a semilocal model in which the vortex {\it
is} stable, and  ask what would happen if $G_2$ were gauged.  Then the vortex
no longer carries any conserved topological charge, and will surely decay.  But
if the $G_2$ gauge coupling is sufficiently weak, one expects the vortex to
remain {\it classically} stable \electroweak.  Thus, the vortex is metastable,
and its decay
is mediated by a magnetic monopole, as we will explain in Section 8.
Similarly, for
$D=3$, there is a metastable string that decays by nucleating a monopole pair.
Because such a metastable string arises in the minimal standard model, for an
(unrealistic \ref\james{M. James, L. Perivolaropoulos, and T. Vachaspati,
``On the Stability of Electroweak Strings,'' Preprint (1992).}) range of values
of $\sin^2\theta_W$ and the Higgs mass,
we refer to
it as an ``electroweak string.''

\newsec{Gauge Hierarchy}

We will now discuss in more detail the decay of metastable defects in a model
with the pattern of {\it gauge} symmetry breaking \aIIi.

\medskip

\noindent {\bf a) Codimension 2}

\noindent {\it i) $D=3$}

We first consider the decay of metastable gauge strings in three
dimensions.  If the string can break due to the nucleation of a
monopole-antimonopole pair, then there will be a nonzero probability of
decay per unit length and time.  A long straight string with string
tension $\mu$ will tunnel to a configuration of the same energy, in
which there is a gap in the string of length $L$, with a monopole at
rest at each end.  The energy cost of producing the pair is $2m$, where
$m$ is the monopole mass; this cost must be balanced by the energy
$\mu L$ saved due to the reduction in the string length.  Thus, the
initial separation of the monopole and antimonopole is $L=2 m/\mu$. The
monopoles subsequently pull apart, consuming the string.
Since the width of the barrier is $L$ and its height is $2m$, we can
crudely estimate that the decay probability will be of the form $P\sim
e^{-B}$, where the WKB tunneling factor is of order $m^2/\mu$ (in units with
$\hbar=1$).

To calculate the semiclassical tunneling factor more precisely, it is most
convenient to use  the Euclidean path integral method \ref\vacdecay{S.
Coleman, {\it Phys. Rev.} {\bf D15} (1977) 2929.}.  The decay is described
by an instanton (or ``bounce''), which
is a solution of Euclidean (imaginary--time, $t=-i\tau$) field equations
approaching the unperturbed string solution at $\tau \rightarrow \pm
\infty$. The origin of $\tau$ can be chosen so that the instanton is
symmetric with respect to $\tau \rightarrow - \tau$. Then the instanton
solution at $\tau = 0$ gives the field configuration at the moment of
nucleation of the monopole--antimonopole pair. The semiclassical decay
probability is
\eqn\eIIi{P \propto e^{-B},}
\noindent where
\eqn\eIIii{B = S - S_0,}
\noindent is the difference between the Euclidean actions of the instanton
($S$) and of the unperturbed string ($S_0$).

In the four--dimensional language, the string axis is represented by a
two--dimensional world sheet, and the monopole center by a one--dimensional
world line. A static straight string oriented along the $z$--axis
corresponds to a planar world sheet, $x=y=0$;
in the instanton solution this world sheet has a ``hole"
which is bounded by the closed world line of the monopole (see Fig.~1).  In
relativistic field theories, the string is invariant with respect to
Lorentz boosts in the $z$--direction, which turn into $z-\tau$ rotations
after the change $t \rightarrow -i\tau$. The bounce solution preserves
this symmetry;
the monopole world line is a circle in the $z-\tau$ plane. In the
rest of this paper we shall assume relativistic invariance, although our
results can be easily generalized to the nonrelativistic case.

The planar string world sheet with a circular hole bounded by the monopole
world line can be thought of as a domain wall bounded by a string in four
spatial dimensions. Indeed, the conditions for the existence of planar and
linear defects in four dimensions are, respectively, $\pi_1 (M) \neq I$ and
$\pi_2 (M) \neq I$. From this point of view, our instanton is a
time--independent solution of ($4 + 1$)--dimensional field equations
describing a planar ``wall" with a circular hole bounded by a ``string" in
a state of unstable equilibrium. If the ``string" radius is decreased, the
hole begins to shrink, and if it is increased, it starts growing.
The bounce solution must have such a negative mode, according to general
arguments \ref\negmode{S. Coleman, {\it Nucl. Phys.} {\bf B298} (1988)
178.}.

If the radius of the monopole world line is much greater than the string
thickness, $R \gg \delta_s$, we can use the thin--string and thin--monopole
approximation, in which the actions for the monopole and for the string are
proportional to the world line length and world sheet area, respectively,
\eqn\eIIiii{S=m\int ds + \mu \int dS_2.}
\noindent For our instantons, the bounce action \eIIii\ is
\eqn\eIIiv{B= 2\pi Rm - \pi R^2\mu.}
This expression is stationary with respect to $R$ for
\eqn\eIIv{R= m/\mu,}
\eqn\eIIvi{B = \pi m^2/\mu.}
\noindent Hence, the initial separation of the monopole--antimonopole pair
is $2m/\mu$, as we anticipated, and the nucleation probability is $P \propto
\exp(-\pi m^2/\mu)$, in agreement with \vildecay.

The thin--defect approximation is justified if \eqn\eIIvii{m/\mu \gg
\delta_s.} \noindent This condition is typically satisfied if the symmetry
breaking scale of monopoles is much greater than that of the strings,
$\eta_1 \gg \eta_2$. Exceptions to this rule can occur if Higgs or gauge
couplings of the model are very small.  For $R \; \roughly{<} \;
\delta_s$, the bounce action depends on the details of the model, and no
simple estimate can be given in the general case.

\medskip

\noindent {\it ii) $D=2$}

In two spatial dimensions, models like \eIiii\ give rise to metastable
vortices.
The vortex can tunnel to a configuration of the same energy, and about
the same core size.  This configuration has a nonzero magnetic field,
but its total magnetic flux is trivial.  Thus, there is nothing to
prevent the configuration from subsequently relaxing to the vacuum.

The instanton in this
case is a monopole--antimonopole pair in unstable equilibrium in
three Euclidean dimensions. The Coulombic attraction between the monopole
and antimonopole is balanced by the tension of the strings pulling them in
opposite directions (see Fig.~2).
%\centerline{\sl Figure 1.~goes here or at the end of the text.}
%\noindent
The theory must have a solution of this form, since we know that
models of
the type \eIiii\ give monopoles connected by strings in three dimensions. The
mid-section of the instanton (surface $\Sigma$ in Fig.~2) gives the field
configuration of the decaying vortex right after the tunneling.

The bounce action can be roughly estimated as
\eqn\eIIxiii{B \sim 2m - {1 \over e^2R} - \mu R,}
\noindent where $m$ is the monopole mass, the second term is the Coulombic
energy of the pair and the last term is the energy of the missing piece of
string. This expression is stationary for
$R \sim e^{-1} \mu^{- 1/2} \sim \delta_s$, where $\delta_s$ is the string
thickness.
Eq.~\eIIxiii\ is only a rough estimate because $R$ is comparable to the
thickness of the string; therefore, the Coulomb interaction between the
monopoles is significantly distorted by the string.
The last two terms in \eIIxiii\ are both of the order
$e^{-1}\mu^{1/2}$ and are negligible compared to the first term if the
symmetry breaking scale of the monopoles is much greater than that of
strings. Hence,
\eqn\eIIxiv{B\approx 2m.}

Recall that the ``mass'' of the monopole actually has the dimensions of
action in $D=2$ (in units with $c=1$).  In order of magnitude it is
$m\sim 4\pi\eta_1/e$, where $\eta_1$ is the expectation value of a scalar
field; $\eta_1$ has the dimensions of $({\rm energy})^{1/2}$, and $e^{-1}$
has the dimensions of $({\rm energy})^{1/2}({\rm length})$, in two
spatial dimensions.

For $\eta_1>>\eta_2$, the effects on the vortex of the physics at energy scale
$\eta_1^2$ can be conveniently incorporated into an effective
Lagrangian \gupta.  Were we to ignore the heavy magnetic monopoles, the low
energy effective theory would have an exact topological conservation law
that would ensure that the vortex is absolutely stable.  But when the
monopole instantons are integrated out, operators are induced in the
effective Lagrangian that violate this conservation law.  Specifically,
an operator appears that annihilates (or creates) a vortex, with a
coefficient that is proportional to $e^{-m}$.  Calculating with this
effective Lagrangian, we again find a vortex decay rate of order
$e^{-2m}$.

\medskip

\noindent {\bf b) Codimension 1}

\noindent {\it i) $D=3$}

The decay of metastable domain walls can be analyzed in a similar way.
If the wall can herniate by nucleating a loop of string, then there will
be a nonzero decay probability per unit time and area \kls.  A planar wall
will tunnel to a configuration of the same energy, with a circular hole
in the wall bounded by the string loop.  The radius of the hole is
chosen so that the energy cost of the string loop matches the energy
saved due to the missing wall.  The string loop appears at rest, and
then expands, consuming the wall.

Again, we compute the tunneling action using the Euclidean path integral
method.  In
the thin--defect approximation, the action is
\eqn\eIIviii{S = \mu \int d S_2 + \sigma \int d S_3,}
\noindent where $\mu$ and $\sigma$ are string and wall tensions,
respectively. In the instanton solution, the wall world membrane is a
three--dimensional hyperplane with a spherical hole bounded by the string
world sheet. The bounce action is then
\eqn\eIIix{B = 4 \pi R^2 \mu - {4 \pi \over 3} R^3 \sigma,}
\noindent which is stationary with respect to $R$ for \kls
\eqn\eIIx{R = 2 \mu /\sigma,}
\eqn\eIIxi{B = {16\pi \mu^3 \over 3\sigma^2}.}

\medskip

\noindent {\it ii) $D=2$}

\medskip

In two spatial dimensions, models like \eIii\ give rise to metastable linear
defects which decay by nucleation of vortex--antivortex pairs. The decay is
described by Eqs.~\eIIiii ---\eIIvi, where now $m$ stands for the vortex mass
and $\mu$ for the tension of the linear defects.

\medskip

\noindent {\it iii) $D=1$}

\medskip

In one spatial dimension, the domain wall becomes a particle, or
``kink.''  Were it not for the existence of vortices, the kink would
carry a conserved topological quantum number, and would be stable.  But
the vortices enable the kink to decay.

The instanton describing the kink decay is a vortex--antivortex pair, in
two Euclidean dimensions, in which the attraction between the vortex and
antivortex is balanced by the tension of linear defects (``walls") attached
to the vortices. An argument similar to the one that led to Eq.~\eIIxiv\ gives
\eqn\eIIxv{B \approx 2 \mu,}
\noindent for the case when the first symmetry breaking scale is much
greater than the second. Here $\mu$ is the vortex action and is given by
the same expression as the string tension in the corresponding
$(3+1)$--dimensional theory.

The kink decays to what might be called a gauged texture, a
configuration that matches the asymptotic behavior of the kink, but
in which the Higgs field is a pure gauge that has a nontrivial winding around
$G/H_1$.  This configuration has a different (1+1-dimensional)
``Chern-Simons number'' than the kink.  The change in the Chern-Simons
number is provided by the vortex instanton.

Again, the violation of the topological conservation law can be
incorporated into an effective Lagrangian.  Integrating out vortices
induces an operator that destroys (or creates) a kink, with a
coefficient proportional to $e^{-\mu}$.

\newsec{Global Hierarchy}

In the case of  global symmetry breaking, the instantons still represent
unstable equilibrium configurations of higher--dimensional defects, but
there is an important difference. Global defects have long-range
interactions mediated by massless Nambu--Goldstone bosons. This leads to an
increase in the height of the potential barrier and to a strong suppression
of the decay. Moreover, since the dominant part of the energy of global
defects is located outside the core, the thin--defect approximation can no
longer be used, and the field configuration of the instanton has to be
studied in some more detail.

\medskip

\noindent {\bf a) Codimension 1}

\medskip

A simple example of a model with metastable domain walls or kinks is
\eqn\eIIIi{L = \vert \partial_\mu \varphi_1\vert^2 + \vert \partial_\mu
\varphi_2 \vert^2 - V(\varphi_1,\;\varphi_2),}
\noindent where $\varphi_1$ and $\varphi_2$ are complex scalar fields and
$$V(\varphi_1,\;\varphi_2) =\lambda_1(\vert \varphi_1\vert^2 -
\eta^2_1)^2 + \lambda_2(\vert \varphi_2\vert^2 - \eta^2_2)^2$$
\eqn\eIIIii{-\lambda_{12}\eta_1 (\varphi_1 \varphi^2_2 + {\rm
h.c.}).\;\;\;\;\;}
This model, for $\eta_1>>\eta_2$, realizes eq.~\eIii\ as a hierarchy of {\it
global} symmetry breaking, with $G=U(1)$.

Without the last term in the potential, the model would have a $U(1) \times
U(1)$ symmetry and $V(\varphi)$ would be minimized by
\eqn\eIIIiii{\varphi_1 = \eta_1 e^{i\theta_1}, \;\; \varphi_2 = \eta_2
e^{i\theta_2},}
\noindent with arbitrary $\theta_1$ and $\theta_2$.
But the last term breaks the symmetry to $U(1)$, and fixes the value of
$\theta_1+2\theta_2$.
To simplify the
analysis, we shall assume that $\lambda_{12}$ is sufficiently small
that it does not affect the magnitudes of the expectation values \eIIIiii.
(Specifically, one needs $\lambda_1 \eta^2_1 \gg \lambda_{12}\eta^2_2, \;
\lambda_2\eta^2_2 \gg \lambda_{12} \eta^2_1$). Then the effective
Lagrangian for the angular variables $\theta_1$ and $\theta_2$ is
\eqn\eIIIiv{L_\theta = \eta^2_1 (\partial_\mu \theta_1)^2 + \eta^2_2
(\partial_\mu \theta_2)^2 + 2\lambda_{12} \eta^2_1 \eta^2_2 \cos(\theta_1 +
2\theta_2).}

We can diagonalize this Lagrangian by
introducing the new variables $\tilde\theta_1$ and $\tilde\theta_2$,
\eqn\eIIIv{\theta_1 = -\tilde\theta_1 + {\eta^2_2 \over
2\eta^2_1}\tilde\theta_2, \;\; \theta_2 = \tilde\theta_2 + {1\over
2}\tilde\theta_1.}
Then, to leading order in $\eta_1 / \eta_2$ we have $\theta_1 + 2\theta_2 =
2\tilde\theta_2$ and
\eqn\eIIIvi{L_\theta = \eta^2_1 (\partial_\mu\tilde\theta_1)^2 +
\eta^2_2(\partial_\mu\tilde\theta_2)^2 + {1\over 2} m^2 \eta^2_2 \cos
2\tilde\theta_2,}
\noindent where $m^2 = 4\lambda_{12}\eta^2_1$. The field $\tilde\theta_1$ is a
massless Nambu--Goldstone field resulting from the
breaking of the global symmetry $\theta_1 \rightarrow \theta_1 + 2
\alpha,\; \theta_2 \rightarrow \theta_2 - \alpha$, while the field
$\tilde\theta_2$ is described by a sine--Gordon Lagrangian. The potential
for $\tilde\theta_2$ is minimized when $\tilde\theta_2 = n\pi$ ($n={\rm
integer}$), and the kink
solution that interpolates between, say, $\tilde\theta_2 = 0$ and
$\tilde\theta_2 = \pi$ is
\eqn\eIIIvii{\tilde\theta_2(x) = 2 \tan^{-1} \exp (mx).}

\noindent The energy of the kink (or tension in the wall) is
\eqn\eIIIviii{\sigma = 4\eta^2_2 m.}

\medskip

\noindent {\it i) $D=1$}

\medskip

In estimating the decay rates, we will start with the case of the kink in one
dimension, and then work our way up to $D=3$.

In the kink solution, $\theta_2$ rotates with $\theta_1$ held fixed.  The kink
can
tunnel to a configuration with the same asymptotic behavior and the same
energy, in which $\theta_2$ and $\theta_1$ rotate together.  This configuration
has gradient energy, but no potential energy, so nothing prevents it from
spreading, and relaxing to a configuration in which $\theta_1$ and $\theta_2$
are nearly constant.

If, after tunneling, the region in which $\theta_1$ twists is of length $L$,
then the gradient energy is of order $\eta_1^2/L$.  This configuration will be
degenerate with the kink for $L\sim \eta_1^2/\sigma$.

To analyze the tunneling more precisely, and compute the tunneling action, we
use the Euclidean path integral method.  The instanton describing the kink
decay is an unstable
defect configuration in two Euclidean dimensions. The two--dimensional theory
with
the Lagrangian \eIIIi, \eIIIii\ has vortex solutions in which the phase
$\theta_1$ changes by $2\pi$ around the vortex. The potential in \eIIIiv\ is
minimized by setting $\theta_2 = - \theta_1/2$, but then the change of
$\theta_2$ around the vortex is only $\pi$, and thus the vortex should be
attached to a linear defect (a ``wall''). The cross--section of the ``wall'' is
identical to the kink \eIIIvii\ and its
tension is given by \eIIIviii. The instanton describing the kink decay consists
of a vortex--antivortex pair held apart by two ``walls.''
Outside the ``walls,''
$\theta_1$ is well approximated by $\theta_1 = \phi_1 + \phi_2$, where $\phi_1$
and $\phi_2$ are azimuthal angles defined in Fig.~3.
The corresponding pattern for $\theta_2$ near the ``walls'' and on the
mid-section $\Sigma$ is sketched in Fig.~4.  The action for this instanton is
\eqn\eIIIix{B = 4 \pi \eta^2_1 \ln (R/\delta_s) - \sigma R;}
\noindent here the first term is twice the energy $\mu$ of a vortex,
\eqn\eIIIx{\mu = \eta^2_1 \int^{\sim R}_{\delta_s} (\nabla \theta_1)^2
\cdot 2\pi r \; dr \approx 2\pi \eta^2_1 \ln(R/\delta_s),}
\noindent $\delta_s \sim \lambda^{-1/2}_1 \eta^{-1}_1$ is the size of the
vortex core and $R$ is the vortex--antivortex separation. Since the
$R$--dependence of $\mu$ is only logarithmic, the result is not sensitive to
whether we choose $R$ or, say, $R/2$ as a cut--off in \eIIIx. The second
term in \eIIIix\ is the energy of the missing ``wall". Eq.~\eIIIix\
is stationary for
\eqn\eIIIxa{R = 4 \pi \eta^2_1/\sigma}
and
\eqn\eIIIxi{B\approx
4\pi\eta_1^2\ln\left({\eta_1^3\lambda_1^{1/2}\over\sigma}\right)
\approx 2\pi \eta^2_1 \ln \left({\lambda_1 \eta^4_1 \over
\lambda_{12} \eta^4_2}\right).}
\noindent  Note that with $\eta_1 \gg \eta_2$, $R$ is much greater than
the ``wall" thickness, $\delta_w \sim m^{-1}$. This justifies the
thin--wall approximation in \eIIIix.

\medskip

\noindent {\it ii) $D=2$}

\medskip

In two spatial dimensions, the model \eIIIi, \eIIIii\ gives rise to metastable
linear defects. The instanton describing their decay consists of a planar
``wall" with a circular hole bounded by a global string in three Euclidean
dimensions. Assuming that $R \gg \delta_w$, the bounce action can be
written as
\eqn\eIIIxii{B \approx 2 \pi R \cdot 2 \pi \eta^2_1 \ln (R/\delta_s) - \pi
R^2 \sigma,}
\noindent which is stationary for
\eqn\eIIIxiii{\sigma R \approx 2 \pi \eta^2_1 \ln (R/\delta_s),}
\noindent and
\eqn\eIIIxiv{B
\approx {4\pi^3\eta_1^4\over
\sigma}\ln^2\left({\eta_1^3\lambda_1^{1/2}\over\sigma}\right)
\approx {\pi^3 \eta^3_1 \over 2 \sqrt{\lambda_{12}} \eta^2_2}
\ln^2 \left({\lambda_1\eta^4_1 \over \lambda_{12} \eta^4_2}\right).}
\noindent  We note that \eIIIxiv\ can be obtained from the thin--defect
result [see Eq.~\eIIvi] using the ``renormalized" global string tension
\eIIIx.

\medskip

\noindent {\it iii) $D=3$}

\medskip

Finally, we consider metastable global defects in three dimensions. The
model \eIIIi, \eIIIii\ has metastable domain walls which decay by nucleation of
circular loops of string. By the same argument as before, the corresponding
tunneling action is given by the thin--defect equations \eIIx, \eIIxi\ with
$\sigma$ from \eIIIviii\ and $\mu$ from \eIIIx,
\eqn\eIIIxx{B \approx {\pi^4 \over 12\lambda_{12}} \left({\eta_1 \over
\eta_2}\right)^4
\ln^3 \left({\lambda_1 \eta^4_1 \over \lambda_{12} \eta^4_2 }\right).}

\medskip

\noindent {\bf b) Codimension 2}

\medskip

As a simple example of a model that contains metastable global strings or
vortices,
consider a model with a spontaneously broken  $SU(2)$ global symmetry, which
has
a scalar
triplet $\vec \varphi_1$ interacting with a doublet $\varphi_2$, via the
potential
$$ V(\vec\varphi_1,\; \varphi_2) = \lambda_1(\vec\varphi_1^2- \eta^2_1)^2 +
\lambda_2
(\varphi^\dagger_2 \varphi_2 - \eta^2_2)^2$$
\eqn\eIIIxv{-\lambda_{12}\eta_1
\vec\varphi_1 \varphi^\dagger_2 \vec\sigma \varphi_2.}
\noindent For $\eta_1>>\eta_2$, the symmetry breaking is that of eq.~\eIiii\
with $G= SU(2)$.

\medskip

\noindent {\it i) $D=2$}

\medskip

In two spatial dimensions, the model \eIIIxv\ contains a metastable vortex.  In
the vortex solution, $\vec\varphi_1$ is essentially a constant, which we may
take to be
\eqn\aIVi{\vec\varphi_1=\pmatrix{0\cr 0\cr \eta_1\cr},}
and $\varphi_2$ has the asymptotic behavior
\eqn\aIVii{\varphi_2(r=\infty,\theta)=\eta_2\pmatrix{e^{i\theta}\cr 0\cr}}
(for $\lambda_{12}>0$).

When we consider the decay of this object, there is a subtlety, namely that the
energy of an isolated vortex is divergent in an infinite volume.  We should
therefore imagine that the vortex is actually a member of a distantly separated
vortex--antivortex pair (or that a suitable infrared cutoff has been imposed in
some other way).  The ``mass'' of the vortex is of order $2\pi
\eta_2^2\ln(R_{\rm
cutoff}/\delta_s)$, where $R_{\rm cutoff}$ is the infrared cutoff and
$\delta_s$ is the size of the vortex core (of order $m_2^{-1}$, where $m_2$ is
the mass of $\varphi_2$).

The vortex can tunnel to a configuration that has the same asymptotic behavior,
but has negligible potential energy.  In this configuration, $\vec\varphi_1$
rotates
inside a region of radius $R$, and the scalar fields lie close to the vacuum
manifold
everywhere.  The gradient energy is then of order $\eta_1^2+\eta_2^2\ln(R_{\rm
cutoff}/R)$ (assuming that $R_{\rm cutoff}>>R$), so this configuration is
degenerate with the vortex for  $\ln(R/\delta_s)\sim \eta_1^2/\eta_2^2$.

If a vortex and antivortex are held at fixed positions, the tunneling of one of
the two is kinematically forbidden unless the distance between them is truly
enormous.  If, say, the antivortex tunnels, the fields will eventually relax to
a
configuration that has a vortex core, but is trivial at spatial infinity, a
configuration with a gradient energy of order $\eta_1^2$.

We turn now to the computation of the tunneling action.
In a three--dimensional space the model \eIIIxv\ has solutions describing
global monopoles attached to global strings, and the instanton describing
the vortex decay consists of a monopole--antimonopole pair held apart by
the string tension. The energy of this configuration diverges not only due
to the infinite length of strings, but also because the energy per unit
length of a global string is logarithmically divergent. However, the bounce
action \eIIii\ can still be expected to be finite, since the monopoles do not
significantly affect the field of the string at distances much greater than
the monopole separation, $R$. Assuming that $R$ is much larger than the
thickness of the string core $\delta_s$, we can write the bounce action as
\eqn\eIIIxvi{B \approx 4\pi \eta^2_1 R - 2\pi \eta^2_2 R \ln (R/\delta_s).}
The first
term in \eIIIxvi\ is the monopole energy, and the second is the energy of the
missing part of the string.

Eq.~\eIIIxvi\ is stationary for
\eqn\eIIIxvii{4\pi \eta^2_1 \approx 2\pi \eta^2_2 \left(\ln {R\over \delta_s} +
1\right),}
\eqn\eIIIxviii{R \approx \delta_s \exp\left({2 \eta^2_1 \over
\eta^2_2}\right),}
\noindent and
\eqn\eIIIxix{B \approx {2\pi \eta^2_2 \delta_s}
\exp \left({2 \eta^2_1 \over \eta^2_2}\right).}
\noindent For $\eta_1 \gg \eta_2$ this action is exponentially large, and
thus the decay of global vortices is very strongly suppressed.
The value of the prefactor in front of the exponential in eq.~\eIIIxix\ should
not be taken seriously; this prefactor is difficult to estimate, because it is
sensitive to terms, subleading in $R$, that have been omitted from
eq.~\eIIIxvi.  To determine the prefactor more accurately, one would have to
solve the field equations for the instanton.

We see that if a vortex and antivortex are held at fixed
positions, with separation $R_{\rm pair}$, the tunneling rate remains finite as
$R_{\rm pair}$ approaches infinity (while the interaction energy of the pair
diverges).  There is of course a competing process, in which the vortex and
antivortex {\it annihilate} due to tunneling.  This is the dominant tunneling
process for $R_{\rm pair}<< R$ (with $R$ as in eq.~\eIIIxviii), but is strongly
suppressed for $R_{\rm pair}>>R$.

\medskip

\noindent {\it ii) $D=3$}

\medskip

Metastable global strings of the model \eIIIxv\ decay by nucleation of
monopole--antimonopole pairs.
Again, the energy per unit length of a global string is infrared divergent, so
we should consider a very long (but finite) string loop, or impose an infrared
cutoff in some other suitable way.

A global monopole and antimonopole attract one another with a force
$4\pi\eta_1^2$ that is independent of their separation.  Thus, it is not so
easy to pay back the energy cost of nucleating the pair by pulling the monopole
and antimonopole apart.  In fact, it is just the logarithmic infrared
divergence in the tension of the global string that makes the tunneling
possible---the energy saved by reducing the string length by $L$ is enhanced by
a factor of $\ln L$ relative to the energy cost of producing a monopole pair
with separation $L$.

The instanton in this case is a planar defect
in four dimensions, with a circular hole bounded by a linear defect
(representing the monopole world line). The tunneling action is
\eqn\eIIIxxi{B \approx 4\pi \eta^2_1\pi R^2 - 2\pi\eta^2_2 \pi R^2 \ln
(R/\delta_s),}
\noindent which is stationary with respect to $R$ for
\eqn\eIIIxxii{R \approx \delta_s \exp\left({2\eta^2_1 \over
\eta^2_2}\right),}
\eqn\eIIIxxiii{B \approx {\pi^2}\eta^2_2 \delta_s^2
\exp \left({4 \eta^2_1 \over  \eta^2_2 }\right).}
\noindent With $\eta_1 \gg \eta_2$, $B$ is typically very large, and the
strings are essentially stable.  (Again, the estimate of the prefactor in front
of the exponential in eq.~\eIIIxxiii\ should not be considered reliable.)

If a finite string loop of radius $R_{\rm loop}$ is held in a fixed position,
then, as in our discussion of vortex decay, there is a competing process in
which the loop annihilates instead of breaking.  But the action for this
process is proportional to $R_{\rm loop}^2$; it is subdominant for
$R_{loop}>>R$ (with $R$ given by eq.~\eIIIxxii).

\newsec{Intrinsic Symmetry Breaking}

As noted in Section 2c, a defect associated with a spontaneously broken global
symmetry may fail to ``survive'' when the symmetry is intrinsically broken by a
small perturbation.  When a defect of codimension $n+1$ does not survive, it
becomes attached to one or more defects of codimension $n$.  If the number of
codimension $n$ defects attached to the codimension $n+1$ defect is exactly
one, then the codimension $n$ defect is metastable, and can decay by nucleating
a codimension $n+1$ defect.  In this section, we illustrate this phenomenon
with a few examples.

\medskip

\noindent {\bf a) Codimension 1}

\medskip

Consider the case, in the notation of eq.~\aIIvii, $G_{\rm
approx}=U(1)\rightarrow H_{\rm approx}=I$, $G_{\rm exact}=Z_N$.  Suppose that,
when we ignore the intrinsic symmetry breaking,  the $U(1)$ symmetry is
spontaneously broken by the condensation of a complex scalar field
\eqn\Vi{\langle \varphi \rangle=\eta e^{i\theta},}
(where the phase $\theta$ is arbitrary).  This model contains a global string,
such that $\theta$ advances by $2\pi$ on a large circle that encloses the
string core.

When we introduce the symmetry breaking perturbation, however, there are $N$
degenerate vacuum states, with
\eqn\Vii{\langle \varphi \rangle=\eta e^{2\pi i k/N},\quad
k=0,1,2,\dots,N-1.}
Hence, the string does not ``survive.''  On a circle surrounding the string
core, $\theta$ will choose to stay close to the vacuum manifold $\theta=2\pi
k/N$, except at isolated points where $\theta$ abruptly jumps from one vacuum
value to the next.  Thus, the string becomes attached to $N$ domain walls
\lazshaf.
(Conceivably, these walls will attract each other, so that in the configuration
of minimal energy $\theta$ jumps by $2\pi$ all in one step.  Then the string is
attached to a single metastable wall.)

The walls have a thickness of order $m_a^{-1}$, where $m_a$ is the mass of the
$U(1)$ pseudo--Goldstone boson (the ``axion'').  The string tension $\mu$ and
the wall tension $\sigma$ are, in order of magnitude
\eqn\Viii{\mu\sim 2\pi\eta^2\ln(1/m_a \delta_s),\quad
\sigma\sim \eta^2m_a,}
where $\delta_s$ is the thickness of the string core.

Note that this precise pattern of symmetry breaking occurs in models that solve
the strong $CP$ problem by the Peccei--Quinn mechanism \nref\pq{R. Peccei and
H.
 Quinn, {\it Phys. Rev. Lett.} {\bf 38} (1977) 1440.}\nref\axion{S. Weinberg,
{\it Phys. Rev. Lett.} {\bf 40} (1978) 223; F. Wilczek, {\it Phys. Rev. Lett.}
{\bf 40} (1982) 279.}\refs{\pq,\axion}, where $U(1)$ is the
Peccei--Quinn symmetry, and the intrinsic symmetry breaking is due to QCD.
\medskip

In the discussion below, we assume that $N=1$, so that a single ``axion domain
wall'' ends on the ``axion string.''

\medskip

\noindent{\it i) $D=1$}

\medskip

In one spatial dimension, this model has a metastable axion ``kink.''  We can
anticipate that this kink, with width of order $m_a^{-1}$, will decay by
tunneling to an ``unwound'' configuration of about the same size.

As in Section 4a.i, the bounce solution is a vortex--antivortex pair in
unstable equilibrium, with the attraction between the pair balanced by the pull
of the kink world lines that are attached to the vortices.  Unlike the
discussion in Section 4a.i, though, the separation of the pair is comparable to
the wall thickness.  Hence, it is difficult to calculate the tunneling action
accurately.

If we repeat our previous analysis (even though it is not well justified here),
we obtain the crude estimates
\eqn\Viv{R\sim 4\pi \eta^2/\sigma\sim m_a^{-1}}
for the vortex separation, and
\eqn\Vv{B\sim 4\pi \eta^2\ln(1/m_a\delta_s)\sim 2\mu}
for the bounce action.

\medskip

\noindent{\it ii) $D=2,3$}

\medskip

In two spatial dimensions, an axion wall decays by nucleating a pair of axion
vortices, and in three spatial dimensions an axion wall decays by nucleating  a
loop of axion string.  These decay processes may be analyzed just as in Section
3b, with $\mu$ and $\sigma$ given by eq.~\Viii.  However, to justify the
thin-defect approximation used there, we must have
\eqn\Vvi{\ln(1/m_a\delta_s)>>1.}

\medskip

\noindent{\bf b) Codimension 2}

\medskip

Now consider the case $G_{\rm approx}=S0(3)\rightarrow H_{\rm approx}=U(1)$,
with $G_{\rm
exact}=I$.  If we ignore the intrinsic symmetry breaking, then this model
contains a texture in two spatial dimensions, or a global monopole
\ref\barvil{M. Barriola and A. Vilenkin, {\it Phys. Rev. Lett.} {\bf 63} (1989)
341.} in three
spatial dimensions.  But when the symmetry breaking perturbation is introduced,
the order parameter has a unique vacuum value.  Hence, the texture collapses to
a point singularity, and the long-range field of the monopole collapses to a
singular line.

If we introduce a short distance cutoff, like a lattice spacing, then the
texture wants to twist in a region with size of order the cutoff.  But on that
scale, there is really no notion of topology that stabilizes the texture, and
there is no reason to expect a metastable defect.

To prevent the texture from shrinking indefinitely, let us introduce into the
action of the model a higher-derivative ``Skyrme term'' \skyrme.  Then there
will be a
classically stable defect whose decay we can analyze semiclassically.

Roughly, the energy of a texture with radius $R$ is
\eqn\Vvii{E_{\rm texture}\sim 4\pi\left( {1\over e_{\rm sk}^2 R^2}
+ \eta^2 + \alpha^2\eta^2 m^2 R^2\right);}
here, the first term is the Skyrme term (with the coupling constant $e_{\rm
sk}^2$ {\it defined} by eq.~\Vvii), the second term is the conventional
gradient
term, and the third term is the potential energy due to the intrinsic symmetry
breaking (where $m$ is mass of the pseudo--Goldstone boson, and $\alpha^2$ is a
constant of order one).  By minimizing with respect to $R$, we find the size of
the texture
\eqn\Vviii{R^2_{\rm texture}\sim {\alpha^{-1}\over e_{\rm sk}\eta~ m}.}
Note that $R_{\rm texture}\to 0$ if we turn off the Skyrme term ($e_{\rm
sk}^2\to\infty$), and that $R_{\rm texture}\to\infty$ if we turn off the
intrinsic symmetry breaking ($m\to 0$).  The mass of the texture is
\eqn\Vix{\mu_{\rm texture}\sim 4\pi\eta^2\left(1+{2\alpha m\over e_{\rm
sk}\eta}\right);}
the second term can be neglected as the Skyrme term or the intrinsic symmetry
breaking turns off.

In eq.~\Vvii\ we have made the assumption that the order parameter is close to
the approximate vacuum manifold $G_{\rm approx}/H_{\rm approx}$.  This
assumption is reasonable only if the energy density inside the texture is small
compared to the energy density of the ``false vacuum'' in which the $G_{\rm
approx}$ symmetry is restored.  The energy density of the false vacuum can be
expressed as $\lambda \eta^4$, where $\lambda$ is a scalar self coupling; thus,
eq.~\Vviii\ and \Vix\ for $\mu_{\rm texture}/R_{\rm
texture}^2 <<\lambda\eta^4$, or
\eqn\Vixa{\left({\lambda\over e_{\rm sk}^2}\right)\left({\alpha m\over e_{\rm
sk}\eta}\right)^{-1}\left(1+{2\alpha m\over e_{\rm sk}\eta}\right)^{-1}>>1.}
If this condition is not satisfied, there is no good reason to expect a
metastable texture to exist.

If we ignore the perturbations, then the global monopole has a core size of
order
\eqn\Vx{R_{\rm core}\sim {1\over \sqrt{\lambda}\eta},}
where $\lambda$ is a scalar self coupling;  this is the linear size of the
region in which the Higgs field departs significantly from its vacuum value.
But the Skyrme term may distort
the core significantly.  If $\lambda/e_{\rm sk}^2>>1$, then the Skyrme term
dominates the gradient energy inside the core, and we find instead
\eqn\Vxi{R_{\rm core}\sim \left({\lambda\over e_{\rm sk}^2}\right)^{1/4}{1\over
\sqrt{\lambda}\eta},\quad \lambda/e_{\rm sk}^2>>1.}
The mass of the
monopole core, in order of magnitude, is
\eqn\Vxiii{M_{\rm core}\sim
\cases{
{4\pi\eta\over \sqrt{\lambda}},& $\lambda/e_{\rm sk}^2 <<1$,\cr
\left({\lambda\over e^2_{\rm sk}}\right)^{3/4}
{4\pi\eta\over \sqrt{\lambda}},& $\lambda/e_{\rm sk}^2 >>1$.\cr}}

The ratio of the texture size to the monopole core size is
\eqn\Vxii
{{R^2_{\rm texture}\over R^2_{\rm core}}\sim
\cases{\left({\lambda\over e^2_{\rm sk}}\right)
\left( {m\over e_{\rm sk}\eta}\right)^{-1},& $\lambda/e_{\rm sk}^2 <<1$,\cr
\left({\lambda\over e^2_{\rm sk}}\right)^{1/2}
\left( {m\over e_{\rm sk}\eta}\right)^{-1},& $\lambda/e_{\rm sk}^2 >>1$.\cr}}
Comparing with the condition eq.~\Vixa, we find that $R_{\rm texture}>>R_{\rm
core}$.  It is reasonable to expect a classically stable texture to exist, and
to treat the intrinsic symmetry breaking as a small perturbation, only if the
texture is large compared to the monopole core.  A monopole attached to a
``texture string'' in three dimensions is illustrated in Fig.~5.

\medskip

\noindent{\it i) $D=2$}

\medskip

In two spatial dimensions, the texture decays by tunneling to an unwound
configuration of about the same size.   The bounce is a monopole--antimonopole
pair in unstable equilibrium, with the attraction of the pair balanced by the
pull of the strings that are attached to the monopoles.  The typical separation
of the pair is comparable to the size of the texture, which makes it difficult
to calculate the tunneling action reliably.

Because $R_{\rm texture}>>R_{\rm core}$, the action of the bounce will be
dominated by the interaction between the monopoles, rather than by the core
action.  In order of magnitude, we expect that
\eqn\Vxiva{B\sim \mu_{\rm texture} R_{\rm texture}\sim
{4\pi\eta\over e_{\rm sk}}\left({\alpha m\over e_{\rm sk}\eta}\right)^{-1/2}
\left(1+{2\alpha m\over e_{\rm sk}\eta}\right).}

\medskip

\noindent{\it ii) $D=3$}

\medskip

In three spatial dimensions, the texture becomes a string, which can decay by
nucleating a monopole pair.  The bounce solution is a planar string world
sheet, punctured by a hole of radius $R$ that is bounded by the world line of a
monopole.  However, $R$ is comparable to the thickness of the string, so we can
not justify the approximation used in Section 3a.i, where we neglected the
string thickness.

Very roughly, we can estimate the order of magnitude of the bounce action as
\eqn\Vxv{B\sim \pi \mu_{\rm string}R_{\rm string}^2\sim
{4\pi\over e_{\rm sk}^2}\left(\alpha m\over e_{\rm sk}\eta\right)^{-1}
\left(1+{2\alpha m\over e_{\rm sk}\eta}\right),}
with $\mu_{\rm string}$ and $R_{\rm string}$ given by eq.~\Vviii\ and \Vix.

\newsec{Decay of Heavy Defects to Light Defects}

We have seen that, in models with a hierarchy of symmetry breakdown, it is
possible for a monopole (or string) that arises at a short distance scale to
become the boundary of a string (or wall) that arises at a longer distance
scale.  In this section, we will comment on another logical possibility.  A
monopole might connect together two distinct types of string, or a string might
connect together two distinct types of wall.  Thus, by nucleating a monopole
pair, a heavy string might decay to a light string.  And by nucleating a
string, a heavy wall might decay to a light wall.  We will illustrate these
possibilities by discussing some particular examples.

\medskip

\noindent {\bf a)  A String Bounding Two Walls}

Consider the sequence of phase transitions
\eqn\VIi{U(1)\times Z_2 \rightarrow Z_2'\times Z_2 \rightarrow Z_2
\rightarrow I,}
where $U(1)$ is a gauge symmetry and $Z_2$ is a global symmetry.  This pattern
of symmetry breaking occurs in a model with  three complex scalar fields
$\phi$, $\psi$, and $\chi$, which carry $U(1)$ charges
\eqn\VIii{Q_\phi=2,\quad Q_\psi= 1, \quad Q_\chi= 1, }
and transform under $Z_2$ as
\eqn\VIiii{Z_2:\quad \phi\to \phi,\quad \psi\to\psi, \quad \chi\to -\chi.}
In this model, $\phi$ condenses at the scale $\eta_1$, breaking $U(1)$ to
$Z_2'$ such that
\eqn\VIiv{Z_2':\quad \phi\to\phi,\quad \psi\to -\psi,\quad \chi\to -\chi.}
Then $\psi$ condenses at $\eta_2<<\eta_1$, breaking $Z_2'$.  Finally $\chi$
condenses at $\eta_3<<\eta_2$, breaking $Z_2$.

The symmetry breaking at scale $\eta_1$ gives rise to a string that eventually
becomes the boundary of both a heavy $\psi$ domain wall at scale $\eta_2$, and
a light $\chi$ domain wall at scale $\eta_3$ (see Fig.~6).  Since the
global $Z_2$ symmetry
is spontaneously broken, there is a stable domain wall in this model, the
$\chi$ wall.  There is also a stable string, which carries twice the $U(1)$
flux of the minimal string that bounds two walls.

\medskip

\noindent {\it i) $D=1$}

In one spatial dimension, this model contains a heavy ($\psi$) kink and a light
($\chi$) kink.
{}From the point of view of an effective field theory that describes physics
well below the scale $\eta_1$, both kinks appear to carry conserved topological
charges, and so should be stable.  But in the underlying theory, there is just
a single topological conservation law that does not forbid the decay of a
$\psi$ kink to a $\chi$ kink.  The decay of the heavy kink is mediated by the
$U(1)$ vortex.  As in the discussion in Section 3b.iii, the action of the
bounce is
\eqn\VIv{B\approx 2\mu,}
where $\mu$ is the vortex action.  (Integrating out the vortex generates an
operator that destroys a $\psi$ kink and creates a $\chi$ kink, with a
coefficient of order $e^{-\mu}$.)

\medskip

\noindent {\it ii) $D=2,3$}

\medskip

In two spatial dimensions, a heavy $\psi$ wall decays to a light $\chi$ wall by
nucleating a pair of vortices, and in three dimensions a $\psi$ wall decays to
a $\chi$ wall by nucleating a loop of string.  These decay processes can be
analyzed as in Section 3b, except that the wall tension $\sigma$ is replaced by
$\sigma_\psi-\sigma_\chi$, the {\it difference} between the  heavy and light
tensions.

\medskip

\noindent {\bf b) A Monopole Bounding Two Strings}

Consider the sequence of phase transitions
\eqn\VIvi{SU(3)\rightarrow U(1)\times U(1)'\rightarrow U(1)\rightarrow Z_2,}
where $U(1)\times U(1)'$  is generated by the diagonal $SU(3)$ generators
\eqn\VIvii{Q={\rm diag}=\left({1\over 2},-{1\over 2},0\right),\quad
Q'={\rm diag}\left({1\over 2},{1\over 2},-1\right),}
and $Z_2$ is generated by $e^{2\pi i Q}=e^{2\pi i Q'}$.  This pattern can occur
in a model with an $SU(3)$ octet that condenses at scale $\eta_1$, a triplet
that condenses at $\eta_2<<\eta_1$, and another octet that condenses at
$\eta_3<<\eta_2$.

When $SU(3)$ breaks to $U(1)\times U(1)'$, there are two conserved magnetic
charges, and so there will be two distinct types of stable magnetic monopole.
The magnetic charges of the stable monopoles are expected to be the minimal
charges $(g_D/2,g_D'/2)$ and $(g_D/2,-g_D'/2)$, where $g_D$ and $g_D'$ are the
Dirac magnetic charges associated with $U(1)$ and $U(1)'$ respectively.  (These
charges satisfy the Dirac quantization condition because $U(1)$ and $U(1)'$
have a nontrivial element in common.)
The two monopoles need not be degenerate, unless there is a charge conjugation
symmetry to enforce the degeneracy.
Monopole solutions with charges $(g_D,0)$ and $(0,g_D')$ may also exist, but
they are likely to be unstable, since they can decay to minimally charged
monopoles, which have lower Coulomb energy.

When the symmetry breaking proceeds further, both
monopoles eventually become attached to two strings---a heavy $U(1)'$ string at
scale $\eta_2$ and a light $U(1)$ string at scale $\eta_3$.
The light string is a stable $Z_2$ string.  The model contains no stable
magnetic monopole.

\medskip

\noindent {i) $D=2$}

\medskip

In two spatial dimensions, this model contains a heavy vortex that carries
$U(1)'$ magnetic flux, and a light vortex that carries $U(1)$ magnetic flux.
If we integrate out the monopoles, and ignore their exponentially small effects
at low energy, then there are two independent conserved vortex numbers, each
taking integer values.

The monopoles break these conservation laws and mediate the decay of a heavy
vortex to a light vortex.
The
decay can be analyzed as in Section 3b.ii, and the bounce action is
\eqn\VIviii{B\approx 2m,}
where $m$ is the monopole action.   A heavy vortex can decay to either a light
vortex or a light antivortex; which decay is favored depends on which of the
two monopole species is lighter.

Since the only exactly conserved vortex quantum number is a $Z_2$ charge, a
pair of light vortices (as opposed to a vortex--antivortex pair) must be able
to annihilate.  The annihilation process involves monopoles of both types, and
has a cross section of order $e^{-2(m_1+ m_2)}$, where $m_{1,2}$ denotes the
monopole action.

\medskip

\noindent {ii) $D=3$}

\medskip

In three spatial dimensions, this model contains a heavy string that can decay
to a light string by nucleating a monopole--antimonopole pair.  The decay can
be analyzed as in Section 3b.i, but with $\mu$ replaced by $\mu_{\rm
heavy}-\mu_{\rm light}$, the {\it difference} between the heavy and light
string tensions.

\newsec{Texture}
As we noted in Section 2d.i, global texture, even if classically stable, can
always decay quantum mechanically.  We discuss a few examples in this section.

\medskip

\noindent {\bf a) $D=1$}

\medskip

To be concrete, we consider a model of a single complex scalar field $\varphi$
with Lagrangian
\eqn\VIIi{L=|\partial_\mu\varphi|^2-\lambda(|\varphi|^2-\eta^2)^
2.}
This model has a spontaneously broken $U(1)$ global symmetry.

In one dimension, a global texture in this model wants to spread out, but we
can stabilize it by imposing an infrared cutoff.  Suppose we take space to be a
circle with circumference $L$.  Then a texture with topological charge $n$  has
the form
\eqn\VIIii{\varphi=e^{2\pi i n x/L},}
and has energy $(2\pi n)^2/L$.  We may take the limit $L\to\infty$ with the
number of twists per unit length $n/L$ held fixed.  In this limit, the texture
has a decay rate per unit time and length that can be computed semiclassically.

Obviously, the texture with $n$ twists will decay to a texture with $n-1$
twists, and we can anticipate that the tunneling will take place in a region
with a size of order $l\equiv L/n$, the length of a single twist.  The
instanton that mediates the decay is the global vortex in two Euclidean
dimensions;  we construct the bounce solution, and compute the tunneling
action, by finding a configuration with a pair of vortices in unstable
equilibrium, with boundary conditions that fix the topological charge per unit
length at $\tau=\pm\infty$.

This problem is actually identical to a problem in two--dimensional
electrodynamics, since the (long--range) interaction between vortices is the
same as the Coulomb interaction.  The boundary conditions place the vortex pair
in a constant background electric field.  In the unstable solution, the
electric force exerted on the vortex by the antivortex is precisely canceled by
the background field.

If the separation between the vortices is $R$, then the action is
\eqn\VIIiii{B\approx 4\pi \eta^2\ln(R/\delta_s)-2\pi \eta^2(R/l).}
The first term is the vortex--antivortex interaction term, and the second term
is due to the interaction of the vortices with the background texture; here,
$\delta_s\sim \lambda^{-1/2}\eta^{-1}$ is the size of the vortex core, and $l$,
again, is the length of a single twist of the texture.  Eq.~\VIIiii\ is
stationary for
\eqn\VIIiv{R\approx 2l}
and
\eqn\VIIv{B\approx 4\pi \eta^2 \ln(l/\delta_s ).}
Eq.~\VIIv\ is valid for $l>>\delta_s$.

(Note that we can do a related calculation for the case of ``gauge texture.''
On a circle of finite length, the number of twists of the Higgs field is a
topological invariant with a gauge invariant meaning, and coincides with the
Chern-Simons number of the gauge field.  But since these configurations are
pure gauge, the states with different winding numbers are degenerate classical
vacuum states, analogous to the ``n-vacua'' of 3+1--dimensional Yang--Mills
theory.  The gauge vortex is the two--dimensional instanton that causes these
classical ground states to mix \ref\vorinst{C. Callan, R. Dashen, and D. Gross,
{\it Phys. Lett.} {\bf 66B} (1977) 375.}.  The mixing splits the degeneracy,
giving rise
to ``$\theta$-vacuum'' states.)

\medskip

\noindent {\bf b) $D=2$}

\medskip

In two spatial dimensions, a global texture has an arbitrary size, and is
marginally stable at the classical level.  Hence, we will not discuss its
semiclassical decay.  The case where the texture is stabilized by suitable
perturbations has already been discussed in Section 5b.

\medskip

\noindent {\bf c) $D=3$}

\medskip

In a model in three spatial dimensions in which a global $G$ symmetry is broken
to $H$, there will be global texture if $\pi_3(G/H)$ is nontrivial \kibble.  A
global
texture wants to shrink, but it can be stabilized if a higher--derivative term
(``Skyrme term'') is introduced into the action of the theory \skyrme.  As was
recently
pointed out by Hindmarsh \hindtex, the Skyrme term can be generated by gauge
boson
exchange if an appropriate {\it subgroup} of $G$ is gauged.

Crudely speaking, the energy of a texture with radius $R$ is of order
\eqn\VIIvi{E_{\rm texture}\sim 2\pi\left({1\over e_{\rm sk}^2 R}+\eta^2
R\right).}
Here, the first term is the Skyrme term, and $e_{\rm sk}^2$ is the
dimensionless Skyrme coupling constant. (In the sort of model considered by
Hindmarsh \hindtex, it is related to a gauge coupling.)  The second term is the
conventional (two-derivative) kinetic term, and $\eta$ is the expectation value
of the order parameter.  Minimizing with respect to $R$, we find the size and
mass of the texture:
\eqn\VIIvii{R_{\rm texture}\sim 1/e_{\rm sk}\eta,\quad\quad M_{\rm texture}\sim
4\pi\eta/e_{\rm sk}.}

In a model that admits global texture, there is always a ``global instanton''
that mediates the decay of the texture.  In four (or more) Euclidean
dimensions, this instanton is attached to a global line defect, which we may
interpret as the world line of the texture.  Inside the core of the instanton,
the spontaneously broken global $G$ symmetry is ``restored.''

If we suppose that the size $R_{\rm core}$ of the instanton core is small
compared to $R_{\rm texture}$, then the action contained inside a sphere of
radius
$R_{\rm texture}$ centered on the core is, in order of magnitude,
\eqn\VIIviii{S_{\rm inst}\sim 2\pi^2\left({1\over e_{\rm sk}^2}\ln (
R_{\rm texture}/R_{\rm core}) + \eta^2(R_{\rm texture}^2 - R_{\rm core}^2)
+\lambda\eta^4 R_{\rm core}^4\right).}
The first term is the Skyrme term, the second is the conventional kinetic term,
and the third term is the potential energy of the core; $\lambda$ is a scalar
self coupling that is {\it defined} by eq.~\VIIviii.  Inside a larger radius
$R>>R_{\rm texture}$, the action is dominated by the linear defect, so we have
$S_{\rm inst}\sim M_{\rm texture}R$.

Assuming that $\lambda/e_{\rm sk}^2>>1$, we find that $S_{\rm inst}$ is
minimized for
\eqn\VIIix{R^4_{\rm core}\sim {e_{\rm sk}^2\over \lambda }~ R_{\rm texture}^4,
}
so that the assumption $R_{\rm core}<<R_{\rm texture}$ is justified, and the
instanton action (cut off at $R\sim R_{\rm texture}$) is
\eqn\VIIx{S_{\rm inst}\sim {\pi^2\over 2e_{\rm sk}^2}\ln (\lambda/e_{\rm
sk}^2).}

When it decays, the texture tunnels to an ``unwound'' configuration with core
size of order $R_{\rm texture}$, which is then free to dissipate.  The bounce
solution that describes this decay is a pair of global instantons in unstable
equilibrium, with the pull of the texture balanced by the
instanton--anti-instanton attraction.  The separation between the instantons in
equilibrium is of order $R_{\rm texture}$, and, for $\lambda/e_{\rm sk}^2>>1$,
the
action of the bounce is
\eqn\VIIxi{B\approx 2S_{\rm inst}\sim {\pi^2\over e_{\rm sk}^2}~\ln
(\lambda/e_{\rm sk}^2).}
The decay is strongly suppressed in the limit $e_{\rm sk}^2\to 0$, where the
texture becomes large.  It is also suppressed, much more weakly, as the
``barrier height'' $\lambda \eta^4$ gets large.

\newsec{Semilocal and Electroweak Strings}

Like the strings that we discussed in Sections 3 and 4, semilocal and
electroweak strings can end on magnetic monopoles, not because of a
symmetry-breaking hierarchy, but for other reasons.  Here we will estimate the
tunneling action for the decay of  semilocal and electroweak vortices, in two
dimensions, and for the breaking of semilocal and electroweak strings, in three
dimensions.  The analysis is similar in spirit to
that described in Sections 3 and 4, but differs in detail.  Actually, our
estimates will be very
crude; to do a better job, one would need to study the interactions of the
monopoles in more detail.

\medskip

\noindent {\bf a)  Semilocal Defects}

\medskip

Recall from the discussion in Section 2d.ii that a semilocal model has
a ``topologically conserved'' magnetic flux (in two dimensions), yet there are
configurations of finite energy in which the flux is spread out over an
arbitrarily large area.  In these configurations, the Higgs field takes values
in the vacuum manifold everywhere, and there is no Higgs field potential
energy.
 If the size $R$ of the configuration is very large, the Coulomb energy of the
magnetic flux can also be neglected; then the only contribution to the energy
is due to Higgs fields gradients.  In two dimensions, gradient energy is scale
invariant, and so remains finite and nonzero as $R\to\infty$.  In the sector
with a single quantum of magnetic flux, let us denote the minimum energy in the
limit $R\to\infty$ by
\eqn\VIIIi{E_\infty=\alpha_\infty \eta^2,}
where $\eta$ is the magnitude of the Higgs field expectation value.  Here
$\alpha_\infty$ is a numerical factor of order one; it depends on the geometry
of the vacuum manifold, but not on any coupling constants or parameters of the
theory.

There are also ``vortex'' configurations, in which the magnetic flux is
confined to a core of finite size.  The characteristic feature of the vortex is
that the stability group of the Higgs field is different at its center than in
the vacuum; thus, the vortex carries Higgs field potential energy.  Let us
suppose that there is a vortex solution to the classical field equations with
energy
\eqn\VIIIii{E_s=\alpha_s\eta^2.}
The structure of the core depends on the detailed dynamics of the theory, so
$\alpha_s$ has a nontrivial dependence on coupling constants.

Now, if $\alpha_s>\alpha_\infty$, then the vortex is not stable.  But it may or
may not be metastable.  In fact, in the one model that has been studied in
detail (the minimal electroweak model in the limit $\sin^2\theta_W=1$), it
turns out that the vortex is either absolutely stable or classically unstable
\nref\semistab{A. Ach\'ucarro, K. Kuijen, L. Perivolaropoulos, and T.
Vachaspati, ``Dynamical Simulations of Semilocal Strings,'' CFA Preprint 3384
(1992).}\refs{\hindprl,\semistab}.
Nevertheless, we will ask what would happen if there is a classically stable
vortex with $\alpha_s>\alpha_\infty$.

To understand the decay of  semilocal strings and vortices, it is important to
recognize that a semilocal string can end on a (global) monopole
\refs{\hindprl,\pressemi}.  In the
notation of Section 2d.ii, the defining property of a semilocal model is
that a closed path that is noncontractible in the gauge orbit $G_1/H_1$ {\it
can} be contracted in the full vacuum manifold $[G_1\times G_2]/H$.  On a large
sphere that surrounds the monopole, a quantum of $G_1/H_1$ magnetic flux enters
through the core of a vortex at, say, the south pole.  The Higgs field
configuration on the sphere excluding the south pole is just a deformation of
the nontrivial loop in $G_1/H_1$ to a point; on each line of constant
latitude, the Higgs field executes a closed path in the vacuum manifold, which
becomes a trivial path at the north pole.  Thus, a quantum of confined magnetic
flux is converted in the core of the monopole to unconfined flux that spreads
and returns to spatial infinity.  The energy of the monopole is infrared
divergent, for the Higgs field gradient energy inside a sphere of radius $R$ is
of order $\alpha_\infty\eta^2 R$ (excluding the energy of the vortex).

To be concrete, let us suppose that the gauge group is $G_1=U(1)$, and is
completely broken.  We denote the gauge coupling by $g'$, in deference to the
analogy with the hypercharge coupling in the standard model.

\medskip

\noindent {\it i) $D=2$}

\medskip

A metastable semilocal vortex will tunnel to a configuration with the same
energy that does not have significant Higgs field potential energy stored in
its core.  If this configuration has radius $R$, we can estimate its energy as
$\alpha_\infty\eta^2+4\pi/g'^2 R^2$, where the first term is due to Higgs field
gradients and the second is due to the magnetic flux.  Equating with the vortex
mass $\alpha_s\eta^2$, we find $R\sim
\sqrt{4\pi}(\alpha_s-\alpha_\infty)^{-1/2}(g'\eta)^{-1}$.

To compute the tunneling action, we construct the bounce.  It consists of a
monopole--antimonopole pair in unstable equilibrium in three--dimensional
Euclidean space (see Fig.~7).  If we assume that $\alpha_s-\alpha_\infty<<1$,
then the pair
will be widely separated, and the interaction ``energy'' can be well
approximated by a linear plus Coulomb potential.  Thus, if $R$ is the
separation, the action is
\eqn\VIIIiii{B\approx 2m_{\rm core}-\alpha_s\eta^2 R+\alpha_\infty \eta^2 R-
4\pi/g'^2 R~;}
the first term is the core action of the monopoles, the second is the action of
the missing string, the third is the linear interaction of the monopoles, and
the fourth is the Coulomb term.  (We have normalized the gauge coupling so that
$4\pi/g'$ is the magnetic flux quantum.)  This expression is stationary for
\eqn\VIIIiv{R^2\approx {4\pi\over (\alpha_s-\alpha_\infty)g'^2\eta^2},}
as we anticipated.  For $\alpha_s-\alpha_\infty$ small, the bounce action is
dominated by the core action of the monopoles,
\eqn\VIIIv{B\approx 2m_{\rm core}}
We expect $m_{\rm core}\sim 4\pi\eta/g'$, for this is the magnetic self--energy
of a monopole with core size of order the vortex width $\delta_s\sim
(g'\eta)^{-1}$.

\medskip

\noindent {\it ii) $D=3$}

\medskip

A metastable semilocal string will decay by nucleating a
monopole--antimonopole pair.  If $\alpha_s-\alpha_\infty$ is small, then
the pair will be sufficiently distantly separated right after the
tunneling that the Coulomb interaction between the monopoles can be
neglected.  If the pair nucleates with separation $L$, then the energy
$\alpha_s\eta^2 L$ saved by removing the string must balance the energy
$2m_{\rm core}+\alpha_\infty\eta^2 L$ of the pair (where $m_{\rm core}$
is the mass of the monopole core).  We conclude that $L\approx 2m_{\rm
core}/(\alpha_s-\alpha_\infty)\eta^2$.

As in our previous calculations of string decay due to monopole pair
nucleation, the bounce solution is a planar string world sheet with a
circular hole of radius $R$, the hole bounded by the world line of the
monopole.  If the Coulomb interaction is neglected, we can calculate $R$
and the tunneling action using a minor modification of the method in
Section 2a.i.  The modification is that the string tension $\mu$ is
replaced by $(\alpha_s-\alpha_\infty)\eta^2$, the {\it difference} between
the string tension and the coefficient in the monopole linear potential,
and $m$ is replaced by the core mass $m_{\rm core}$.  Then, from
eq.~\eIIv\ and \eIIvi, we find
\eqn\VIIIvi{R\approx {m_{\rm core}\over (\alpha_s-\alpha_\infty)\eta^2}}
and
\eqn\VIIIvii{B\approx {\pi m_{\rm core}^2\over
(\alpha_s-\alpha_\infty)\eta^2}.}

\medskip

\noindent {\bf b) Electroweak Defects}

\medskip

Now consider the case of a semilocal model with
$\alpha_s<\alpha_\infty$, so that a vortex is {\it stable}. Let us ask
what would happen if we were to gauge the global $G_2$ symmetry.  To be
concrete, consider an interesting example---the standard electroweak
model with gauge group $SU(2)_L\times U(1)_Y$ and a Higgs doublet.  If
we turn off the $SU(2)_L$ gauge coupling $g$, this becomes a semilocal
model;  the gauge group $U(1)_Y$ is broken, but a noncontractible loop
in the gauge orbit is contractible in the full vacuum manifold.  The
vortex turns out to be stable, in this limit, if the Higgs mass is less
than the $Z^0$ mass \refs{\hindprl,\semistab}.

For $g\ne 0$, there is no longer a topological conservation law, and the
vortex is no longer absolutely stable.  Only a finite energy barrier separates
a
vortex with heavy $Z^0$ magnetic flux from a configuration with massless $A$
magnetic flux that is free to spread out.  Correspondingly (as Nambu
\ref\nambu{Y. Nambu, {\it Nucl. Phys.} {\bf B130} (1977) 505.}
observed long ago), a $Z^0$ string can end on an {\it electromagnetic}
monopole (see Fig.~8).  The quantity of $Z^0$ flux trapped in the string is
\refs{\nambu,\pressemi}
\eqn\VIIIviii{\Phi_Z=4\pi\sin\theta/g',}
and the quantity of $A$ flux emanating from the monopole is
\eqn\VIIIviii{\Phi_A=4\pi\sin\theta/g}
(where $\tan\theta\equiv g'/g$).

In the limit $g<<g'$ (or $\sin\theta\approx 1$), the monopole has a core
size that is large compared to the thickness of the string $\delta_s\sim
m_Z^{-1}$ (where $m_Z$ is the $Z^0$ mass).  Deep inside the core, it
resembles the semilocal monopole, with spreading $Z^0$ flux.  At a
radius $R_{\rm core}$, the $Z^0$ flux is converted to $A$ flux.  The
core radius is determined by the competition between the linearly
divergent energy of the ``global'' monopole and the magnetic self
energy.  Roughly, the core energy is
\eqn\VIIIix{E_{\rm core}\sim \alpha_\infty\eta^2 R_{\rm core}
+ {2\pi\sin^2\theta\over g'^2}(m_Z-R_{\rm core}^{-1})
+{2\pi\sin^2\theta\over g^2} R_{\rm core}^{-1},}
where the second term is due to the $Z^0$ flux and the third term is due
to the $A$ flux.  By minimizing with respect to $R_{\rm core}$, we find
\eqn\VIIIx{R_{\rm core}^2\approx {-2\pi\cos 2\theta\over \alpha_\infty
g^2\eta^2}\approx {2\pi\over \alpha_\infty g^2\eta^2}}
(the second equality following from our assumption $g<< g'$), and
\eqn\VIIIxi{E_{\rm core}\approx \sqrt{4\pi\alpha_\infty}~~{\eta\over g}}.

\medskip

\noindent {\it i) $D=2$}

\medskip

In two dimensions, an electroweak vortex decays by tunneling to a configuration
in which its $Z^0$ flux has been converted to $A$ flux.  In the limit $g<<g'$
the flux $\Phi_A$ given by eq.~\VIIIviii\ is much larger than $\Phi_Z$.
Therefore, $A$ flux is energetically very costly; to be degenerate with the
vortex, the configuration after tunneling must be very large.

The bounce solution is a monopole--antimonopole pair in unstable equilibrium in
three Euclidean dimensions, with the Coulomb attraction of the pair compensated
by the tension in the electroweak strings.
For typical values of the parameters, the separation $R$ of the pair will be
comparable to the size $R_{\rm core}$
of the monopole, which makes it difficult to estimate the tunneling action
reliably.
But if we assume that $R>>R_{\rm core}$, then the action of the configuration
is
\eqn\VIIIxii{B\approx 2E_{\rm core} -\alpha_s\eta^2 R
- {4\pi \sin^2\theta \over g^2 R},}
which is stationary for
\eqn\VIIIxiii{R^2\approx {4\pi \sin^2\theta\over \alpha _s g^2\eta^2}\approx
{4\pi \over \alpha_s g^2\eta^2}}
(assuming $\sin^2\theta \approx 1$.)  Thus, our assumption $R>>R_{\rm core}$ is
justified under the (not very physical) condition
\eqn\VIIIxiv{\alpha_s/\alpha_\infty << 1.}
The tunneling action is
\eqn\VIIIxv{B\approx 2 E_{\rm core}-\sqrt{4\pi \alpha_s}~\sin\theta~
{\eta\over g}\approx 2E_{\rm core};}
it is dominated by the core action for $\alpha_s<<\alpha_\infty$.

\medskip

\noindent{\it ii) $D=3$}

\medskip

An electroweak string decays by nucleating a monopole--antimonopole pair.
Typically, the string will tunnel to a configuration in which the separation of
the pair is comparable to the monopole core size, but if we assume that
$\alpha_s<<\alpha_\infty$, then the core is sufficiently small that the
analysis of Section 3a.i applies, with $m=E_{\rm core}$ and
$\mu=\alpha_s\eta^2$.  Thus, the monopole world line has radius
\eqn\VIIIxvi{R\approx E_{\rm core}/\alpha_s\eta^2
\approx {\sqrt{4\pi\alpha_\infty}\over \alpha_s g\eta}}
(for $\sin^2\theta\sim 1$), and the bounce action is
\eqn\VIIIxvii{B\approx \pi E_{\rm core}^2/\alpha_s \eta^2\approx
{4\pi^2\over g^2}{\alpha_\infty\over \alpha_s}.}
So, of course, the decay is heavily suppressed as $g^2\to 0$.

\newsec{Concluding Remarks}

\noindent {\it i) Summary}

\medskip

In this paper, we have studied the decay of metastable defects arising from
symmetry
breaking in relativistic field theories.  The decay occurs through quantum
tunneling; for example, strings decay by nucleation of monopole-antimonopole
pairs, and domain walls decay by nucleation of circular holes bounded by
strings.  We also studied the decay of defects in one and two-dimensional
systems and the decay of non-topological defects, such as global texture and
semilocal and electroweak strings.

The decay probability is determined by an instanton which can be found by
solving Euclideanized field equations with appropriate boundary conditions.
The problem is greatly simplified when the dimensions of the instanton are much
greater than the size of the defect core, so that the thin defect approximation
can
be used.  We assumed the validity of this approximation in most of our
calculations.

We have estimated the instanton action for the decay of various defects in
$D$=1, 2, or 3 spatial dimensions.  We found that the decay rate of
defects arising from a global symmetry breaking is strongly suppressed compared
to the decay rate in the corresponding gauge theory.  In particular, the
instanton action
for the decay of a global string is exponentially large (see eq.~\eIIIxxiii).
But
even in gauge theories, the tunneling action is large and the decay rate is
correspondingly small in the case where two different symmetry breaking scales
are widely separated.

In conclusion, we would like to comment on some possible applications and
extensions
of our results.

\medskip

\noindent {\it ii) Thin Defect Approximation}

\medskip

The thin defect approximation is similar to the thin wall approximation in the
vacuum decay problem \vacdecay.  This latter approximation is in bad
repute, since it is known to apply only to a very limited class of potentials
\ref\hiscock{D. A. Samuel and W. A. Hiscock, {\it Phys. Lett.} {\bf B261}
(1991) 251.}.
On the other hand, the thin-defect approximation applies if the symmetry
breaking scales for the two types of defects involved in the decay are
substantially different.  This is a typical situation in elementary particle
theories, and thus we expect our results to have a reasonably wide range of
validity.

\medskip

\noindent {\it iii) ``Embedded'' Defects}

\medskip

Let us briefly consider the behavior of the decay rate when the two symmetry
breaking scales are close together.  Though the thin--defect approximation does
not apply in this case, we can roughly estimate the tunneling action by using
our thin--defect formulas.  For example, consider the breaking of a string due
to monopole nucleation.  The monopole mass can be crudely estimated as
\eqn\IXa{m\sim 4\pi\eta_1/e,}
where $\eta_1$ is the higher symmetry breaking scale and $e$ is the gauge
coupling, and the string tension is roughly
\eqn\IXb{\mu\sim 2\pi\eta_2^2,}
where $\eta_2$ is the lower symmetry breaking scale.  Then the tunneling action
eq.~\eIIvi\ becomes
\eqn\IXc{B\approx \pi m^2/\mu\sim {8\pi^2\over e^2}\left({
\eta_1\over \eta_2}\right)^2.}
Naturally, the behavior of this expression as $\eta_1$ approaches $\eta_2$
(aside from the numerical factor $8\pi^2$, which should not be taken too
seriously anyway), could be determined by dimensional analysis, as $e^{-2}$ has
the dimensions of $\hbar$.

There is no guarantee that a {\it classically stable} string solution will
continue to exist as the two symmetry breaking scales approach each other.
However, if the Higgs potential obeys suitable conditions, one can argue that
there {\it is} a static string solution for $\eta_1=\eta_2$, although it may be
unstable.  This is a special case of the ``embedded'' string recently described
by Vachaspati and Barriola \ref\vachbarr{T. Vachaspati and M. Barriola, ``A New
Class of Defects,'' Preprint (1992).}.  Eq.~\IXc\ suggests that, if an embedded
string is classically stable, its quantum--mechanical decay rate is likely to
be small, in a weakly coupled theory.  Of course, it is a fact of life that if
our semiclassical (small $\hbar$) approximation is reliable, then the tunneling
action is large and the decay rate is small.

We can do a similar estimate of the rate for the decay of a domain wall due to
nucleation of a string loop.  Taking the string tension to be
\eqn\IXd{\mu\sim 2\pi\eta_1^2}
and the wall tension to be
\eqn\IXe{\sigma\sim 2\sqrt{\lambda}\eta_2^2,}
where $\lambda$ is a scalar self coupling, eq.~\eIIxi\ becomes
\eqn\IXf{B\approx {16\pi\mu^3\over 3\sigma^2}\sim
 {32\pi^4\over 3 \lambda}\left({\eta_1\over \eta_2}\right)^6.}
If there is a classically stable ``embedded'' domain wall for $\eta_1=\eta_2$,
it will be long--lived at weak coupling.

\medskip

\noindent {\it iv) Cosmological Applications}

\medskip

The very small decay rates for metastable defects do not necessarily mean that
such decay processes are observationally irrelevant.  Consider for example
metastable strings formed in the sequence of cosmological phase transitions
\eIiii.  The first phase transition gives rise to monopoles and the second to
strings which connect monopole--antimonopole pairs.
Even if we disregard the breaking of the string due to the quantum--mechanical
decay process,
such hybrid defects
typically disappear long before the present epoch \vildecay.  The string
energy is dissipated by friction and by radiation of gauge quanta and of
gravitational waves.  As the strings get shorter, the monopoles are pulled
closer together;  they eventually capture one another into Coulombic bound
states and annihilate.  In this scenario, the demise of the monopole--string
system is so rapid that the slow monopole nucleation process
has very little effect.

But the evolution can be quite different if there is a period of inflation
between
the two phase transitions in \eIiii.  In this case the monopoles can be
diluted beyond the present horizon, and the evolution of strings will initially
be similar to that of topologically stable strings.  However, at some point the
string decay by monopole pair nucleation will become important.  To estimate
the time $t_*$ when this happens, we write the nucleation probability per unit
string length per unit time as
\eqn\IXi{ P=A\mu \exp (-\pi m^2 /\mu) ,}
\noindent
where $A$ is a dimensionless coefficient and we have used \eIIvi\ for the
tunneling action.  The strings stretching across the horizon at cosmic time $t$
have length of order $t$; so the condition for approximately one pair per
horizon volume to nucleate on a given string by cosmic time $t_*$ is $ P{t_*}^2
\sim 1$, or
\eqn\IXii{t_* \sim (A\mu)^{-1/2} \exp (\pi m^2 /2\mu) .}
\noindent
With a grand unification symmetry breaking scale for strings, $\mu^{1/2} \sim
10^{16}$ GeV, and assuming that $A$ is not very different from 1, this time is
smaller than the present age of the universe if $m^2 /\mu \leq 85$.

At $t \sim t_*$ the strings are cut into pieces of length $\sim t_*$ with
monopoles at the ends.  For large values of $t_*$ it may take a long time to
dissipate the string energy, and oscillating string pieces may still be flying
somewhere in the universe.  As they are pulled by the strings, the monopoles
are accelerated to energies comparable to the energy of the string, $ E
\sim \mu t_*$.  If $t_*$ is close to the present time, then, for
grand--unification strings this energy corresponds to a mass  of order
$10^{17}$
solar masses.  A monopole will move ultrarelativistically, with its
gravitational field concentrated in the transverse plane and the gravitational
force decreasing as $r^{-1}$ with the distance from the monopole.
Gravitational effects of such supermassive relativistic objects can be quite
significant.

\medskip

\noindent {\it v) Nonzero Temperatures}

\medskip

Our results can be easily extended to the case of defect decay at a nonzero
temperature, T.  The Euclidean path integral in this case is taken over field
configurations periodic in imaginary time $\tau$ with a period $\Delta \tau =
T^{-1}$, and the instanton solutions should have the same periodicity.  At
sufficiently low temperatures, the finite--$T$ instanton is simply the
zero--$T$ instanton periodically repeated along the
$\tau$--axis\nref\affleck{I. Affleck, {\it Phys. Rev. Lett.} {\bf 46} (1981)
388.}
\nref\lindecay{A. D. Linde, {\it Nucl. Phys.} {\bf B216} (1983) 421.}
\refs{\affleck,\lindecay}.  The effects
of periodicity become important when $T$ becomes comparable to the inverse size
of the instanton, $R^{-1}$.  At still higher temperatures, the decay of the
defect tends to be dominated by thermal fluctuations rather than quantum
fluctuations.  The interesting and somewhat unexpected behavior
of instantons at higher temperatures will be discussed in a separate paper
\ref\jaume{J. Garriga and A. Vilenkin, to be published}.

\medskip

\noindent {\it vi) Condensed Matter Applications}

\medskip

Condensed matter systems exhibit a fascinating variety of defects,
many of which are metastable, at least in principle.  For example, superfluid
${}^3$He-$B$ contains domain walls that can terminate on strings
\ref\salomaa{M. M. Salomaa
and G. E. Volovik, {\it Phys.Rev.} {\bf B37} (1988) 9298.}, and both nematic
liquid crystals \ref\williams{C. Williams, P. Pieranski and
P. E. Cladis, {\it Phys. Rev. Lett.} {\bf 29} (1972) 90.} and ${}^3$He-$A$
\ref\anderson{P. W. Anderson and G. Toulouse, {\it Phys. Rev. Lett.} {\bf 38}
(1977) 403.} contain strings that can terminate on monopoles.  In suitable
materials, the decay of walls due to string nucleation and of strings due to
monopole nucleation may occur at observable rates.  Also, liquid crystals and
${}^3$He can contain textures that may decay either quantum mechanically
or due to thermal fluctuations.

In these cases, the simplifying assumptions of Lorentz invariance and the
thin--defect approximation may not apply.  But the instantons corresponding to
the decay processes have the same general structure as the instantons that we
have constructed, and the decay rates can be estimated using the methods
described here.

\bigbreak\bigskip\bigskip\centerline{{\bf Acknowledgements}}\nobreak
This work was supported in part by DOE contract DE-AC03-81-ER40050 and by NSF
contract 8605578.  A. V. gratefully acknowledges the hospitality of Caltech,
where he visited as a Sherman Fairchild Distinguished Scholar.

\listrefs

%\listfigs   %(if necessary)

\figures

\fig{1}{Instanton describing the decay of a metastable string, with
one dimension suppressed.  The field configuration in the hyperplane $\Sigma$
corresponds to the moment of nucleation of the monopole--antimonopole pair.}

\fig{2}{Instanton for the decay of a metastable vortex is a
monopole--antimonopole pair in unstable equilibrium inside a string.  The
magnetic field pattern in a plane containing the monopoles is shown.  Shading
indicates the area containing most of the magnetic flux in the
string cross-section.}

\fig{3}{Instanton for the decay of a global kink.  The angles $\phi_1$ and
$\phi_2$ are defined as shown, with branch cuts at the centers of the
``walls.''
The shaded region is the interior of the kink, where $\tilde\theta_2 \approx
\theta_2 + \theta_1 /2$ is substantially different from $0$ or $\pi$.}

\fig{4}{Instanton for the decay of a global kink.  The directions of arrows
indicate the value of $\theta_2$.}

\fig{5}{Global monopole attached to a ``texture string.''  The direction of
the Higgs triplet is shown by arrows.  The field configuration in the
asymptotic
region is trivial.}

\fig{6}{Cross-section of a string attached to light and heavy walls.}

\fig{7}{Instanton for semilocal vortex decay.  The magnetic field pattern in a
plane containing the monopoles is shown.}

\fig{8}{Electroweak string terminating on an electroweak monopole.  The
magnetic field pattern on a slice through the string and monopole is shown.
The monopole converts the (confined) $Z^0$ flux to (unconfined) electromagnetic
flux.}

\bye